\documentclass[10pt,letterpaper]{article}
\usepackage[top=0.85in,left=2.75in,footskip=0.75in]{geometry}

\usepackage{amsmath,amssymb}

\usepackage{changepage}

\usepackage[utf8x]{inputenc}

\usepackage{textcomp,marvosym}

\usepackage{cite}

\usepackage{nameref,hyperref}

\usepackage{microtype}
\DisableLigatures[f]{encoding = *, family = * }

\usepackage[table,dvipsnames]{xcolor}

\usepackage{array}

\newcolumntype{+}{!{\vrule width 2pt}}

\newlength\savedwidth

\raggedright
\usepackage[aboveskip=1pt,labelfont=bf,labelsep=period,justification=raggedright,singlelinecheck=off]{caption}

\makeatletter
\renewcommand{\@biblabel}[1]{\quad#1.}
\makeatother

\usepackage{stmaryrd, bm, amsthm, booktabs, xspace, makecell}
\usepackage{longfbox}

\usepackage{lastpage,fancyhdr,graphicx}
\usepackage{epstopdf}
\fancyheadoffset[L]{2.25in}
\fancyfootoffset[L]{2.25in}
\graphicspath{{./img/}}

\usepackage{cleveref}

\definecolor{BrewerRed}{RGB}{228,26,28}
\definecolor{BrewerGreen}{RGB}{77,175,74}
\definecolor{BrewerBlue}{RGB}{55,126,184}
\definecolor{BrewerPurple}{RGB}{152,78,163}
\definecolor{BrewerOrange}{RGB}{255,127,0}
\definecolor{BrewerYellow}{RGB}{215,215,41}
\definecolor{BrewerGray}{RGB}{150,150,150}

\newcommand\bX{\ensuremath{\bm X}\xspace}

\newcommand\PI[2]{\ensuremath{I_{\partial}^{#1\rightarrow #2}}\xspace}
\newcommand\IR[2]{\ensuremath{I_{\cap}^{#1\rightarrow #2}}\xspace}
\newcommand\wmsphi{\ensuremath{\Phi^{\mathrm{WMS}}}\xspace}

\newcommand{\phiid}{\ensuremath{\Phi\mathrm{ID}}\xspace}
\newcommand\phiR{\ensuremath{\Phi^{\mathrm{R}}}\xspace}

\newcommand\rtr{\ensuremath{\texttt{Red}\!\rightarrow\!\texttt{Red}}\xspace}
\newcommand\rtx{\ensuremath{\texttt{Red}\!\rightarrow\!\texttt{Un}^1}\xspace}
\newcommand\rty{\ensuremath{\texttt{Red}\!\rightarrow\!\texttt{Un}^2}\xspace}
\newcommand\rts{\ensuremath{\texttt{Red}\!\rightarrow\!\texttt{Syn}}\xspace}
\newcommand\xtr{\ensuremath{\texttt{Un}^1\!\rightarrow\!\texttt{Red}}\xspace}
\newcommand\xtx{\ensuremath{\texttt{Un}^1\!\rightarrow\!\texttt{Un}^1}\xspace}
\newcommand\xty{\ensuremath{\texttt{Un}^1\!\rightarrow\!\texttt{Un}^2}\xspace}
\newcommand\xts{\ensuremath{\texttt{Un}^1\!\rightarrow\!\texttt{Syn}}\xspace}
\newcommand\ytr{\ensuremath{\texttt{Un}^2\!\rightarrow\!\texttt{Red}}\xspace}
\newcommand\ytx{\ensuremath{\texttt{Un}^2\!\rightarrow\!\texttt{Un}^1}\xspace}
\newcommand\yty{\ensuremath{\texttt{Un}^2\!\rightarrow\!\texttt{Un}^2}\xspace}
\newcommand\yts{\ensuremath{\texttt{Un}^2\!\rightarrow\!\texttt{Syn}}\xspace}
\newcommand\str{\ensuremath{\texttt{Syn}\!\rightarrow\!\texttt{Red}}\xspace}
\newcommand\stx{\ensuremath{\texttt{Syn}\!\rightarrow\!\texttt{Un}^1}\xspace}
\newcommand\sty{\ensuremath{\texttt{Syn}\!\rightarrow\!\texttt{Un}^2}\xspace}
\newcommand\sts{\ensuremath{\texttt{Syn}\!\rightarrow\!\texttt{Syn}}\xspace}

\newtheorem{theorem}{Theorem}%

\newtheorem{proposition}[theorem]{Proposition}

\newtheorem{definition}{Definition}

\newcommand\subref[2]{\hyperref[#1]{\ref*{#1}#2}}

\begin{document}
\vspace*{0.2in}

\begin{flushleft}
{\Large
\textbf{
Towards an extended taxonomy of information dynamics via Integrated Information Decomposition}
}
\newline
\\
Pedro A.M. Mediano\textsuperscript{1*},
Fernando E. Rosas\textsuperscript{2,3,4*},
Andrea I. Luppi\textsuperscript{5,6,7},
\mbox{Robin L. Carhart-Harris}\textsuperscript{2},
Daniel Bor\textsuperscript{1},
Anil K. Seth\textsuperscript{8},
Adam B. Barrett\textsuperscript{8,9}
\\
\bigskip
\textbf{1} Department of Psychology, University of Cambridge, Cambridge, UK
\\
\textbf{2} Centre for Psychedelic Research, Imperial College London, London, UK
\\
\textbf{3} Data Science Institute, Imperial College London, London, UK
\\
\textbf{4} Centre for Complexity Science, Imperial College London, London, UK
\\
\textbf{5} University Division of Anaesthesia, University of Cambridge, Cambridge, UK\\
\textbf{6} Department of Clinical Neurosciences, University of Cambridge, Cambridge, UK
\\
\textbf{7} Leverhulme Centre for the Future of Intelligence, University of Cambridge, Cambridge, UK
\\
\textbf{8} Sackler Center for Consciousness Science, University of Sussex, Brighton, UK
\\
\textbf{9} The Data Intensive Science Centre, University of Sussex, Brighton, UK
\\
\bigskip

* These authors contributed equally to this work.\\
\hphantom{*} Correspondence: pam83@cam.ac.uk; f.rosas@imperial.ac.uk

\end{flushleft}
\section*{Abstract}

Complex systems, from the human brain to the global economy, are made of
multiple elements that interact in such ways that the behaviour of the `whole'
often seems to be more than what is readily explainable in terms of the `sum of
the parts.' Our ability to understand and control these systems remains
limited, one reason being that we still don't know how best to describe -- and
quantify -- the higher-order dynamical interactions that characterise their
complexity. To address this limitation, we combine principles from the theories
of Information Decomposition and Integrated Information into what we call
\textit{Integrated Information Decomposition}, or $\Phi$ID. $\Phi$ID provides a
comprehensive framework to reason about, evaluate, and understand the
information dynamics of complex multivariate systems. $\Phi$ID reveals the
existence of previously unreported modes of collective information flow,
providing tools to express well-known measures of information transfer and
dynamical complexity as aggregates of these modes. Via computational and
empirical examples, we demonstrate that $\Phi$ID extends our explanatory power
beyond traditional causal discovery methods -- with profound implications for
the study of complex systems across disciplines.


\section*{Introduction}

How can we best characterise the plethora of dynamical phenomena that can
emerge in a system of interacting components? Progress on this question seems
critical to support advances in our ability to understand, engineer, and
control complex systems such as the central nervous
system~\cite{kelso1995dynamic}, the global climate~\cite{runge2019inferring},
macroeconomics \cite{Dosi:2019}, and many others. The predominant approach for
analysing such systems is in terms of cause-effect pairs, seeking to link each
cause to its individual effect via a `causal arrow' (see
e.g.~\cite{bressler2011wiener,pearl2018book}). However, such approaches have an
important limitation that is rarely acknowledged: they neglect higher-order
relationships that cannot be expressed in terms of such arrows, which --
nonetheless -- have been shown to be ubiquitous in complex systems as varied as
genetic networks~\cite{chan2017gene}, financial
markets~\cite{scagliarini2020synergistic}, baroque music~\cite{Rosas2019}, and
the human
brain~\cite{wibral2015partial,luppi2020synergisticCognition,luppi2020synergisticConsciousness}.

As an alternative approach, various one-dimensional metrics have been proposed
to assess the dynamical complexity of processes
(e.g.~\cite{grassberger1986toward,crutchfield2003regularities}). An interesting
case of this is found in the neuroscience literature, where it has been
proposed that a key feature of the neural dynamics underpinning advanced
cognition, flexible behaviour, and even the presence of consciousness, can be
captured by a single number that accounts for the brain's ability to `integrate
information.' There have been several attempts to operationalise this notion,
including the various $\Phi$ measures in Integrated Information Theory
(IIT)~\cite{Tononi1994,Balduzzi2008,Oizumi2014} and Causal Density
(CD)~\cite{Seth2011}; however, these measures have been shown to behave
inconsistently~\cite{van2003neural,Mediano2019,Rosas2019}, making empirical
applications difficult to interpret.

Here we argue that the two research programs described above are limited
because they are rooted in the intuition that information can only be
\textit{transferred} or \textit{stored} between parts of a system. However, it
has recently become apparent that \textit{high-order interactions} that go
beyond transfer and storage can be captured via the framework of Partial
Information Decomposition (PID)~\cite{james2016information}, which demonstrates
how the information that two or more sources provide about a given target can
be decomposed into redundant, unique, and synergistic
components~\cite{Williams2010}. Specifically, redundancy refers to information
held simultaneously by both sources, unique information is that held by one
source but not the other, and synergy is the information conveyed by both
sources together but none of them in isolation.
PID has been usefully applied to systems such as cellular
automata~\cite{finn2018quantifying,rosas2018information}, artificial neural
networks~\cite{beer2015information,tax2017partial}, socioeconomic
data~\cite{varley2021intersectional}, and gene
interactions~\cite{cang2020inferring}.
However, the information taxonomy introduced by PID is only valid in scenarios
with a single target variable, being unable to discriminate between different
ways in which two or more target variables can be affected collectively. This
important limitation prevents PID from providing an encompassing view of the
dynamics of complex systems, where the past of multiple variables affects the
future of multiple variables.

In this paper we introduce the \emph{Integrated Information Decomposition}
framework ($\Phi$ID), which combines principles from the theories of
Information Decomposition and Integrated Information to overcome PID's critical
limitation and enables a complete information decomposition on groups of time
series. The $\Phi$ID framework introduces a novel information taxonomy,
revealing the existence of modes of information dynamics that have not been
previously reported. Furthermore, it allows us to show precisely how measures
of transfer entropy and integrated information are aggregates of several
qualitatively distinct modes. As proof of concept, we use two example datasets
(simultaneous recordings of heart rate and respiratory volume in healthy
subjects; and a neurobiologically realistic simulation of whole-brain activity)
to show that the high-order effects discussed here are not merely theoretical
speculations, but can have substantial effects on real-world analyses and their
interpretation.

\section*{Results}

\subsection*{Integrated information decomposition: $\Phi$ID}

\subsubsection*{Decomposing multivariate information}

Consider a system composed of two interdependent elements that co-evolve over
time. If the system's future state depends only on the preceding state (i.e. if
its dynamics are Markovian), then the total amount of information carried from
past to future is known as the \textit{time-delayed mutual
information}\footnote{In non-Markovian systems the corresponding quantity is
known as \textit{excess entropy}~\cite{crutchfield2003regularities}.
Information decomposition in non-Markovian systems will be covered in a
subsequent publication.}~\cite{Mediano2019}, and can be quantified as the
mutual information between past and future states of the system:
\begin{equation}\label{eq:excess_entropy_markov}
\mathrm{TDMI} = 
I( \bX_t ; \bX_{t+1} ) =
I(X^1_t,X^2_t \; ; \; X^1_{t+1},X^2_{t+1}) ~ .
\end{equation}
Above, the superscripts $1$ and $2$ refer to the two elements of the system,
and $\bX_t= (X^1_t,X^2_t)$ is a shorthand notation for the system's state at
time $t$.

One can analyse the total information flow using the \textit{Partial
Information Decomposition} (PID) framework, which decomposes the mutual
information between multiple sources and a target variable into unique
(\texttt{Un}), redundant (\texttt{Red}), and synergistic (\texttt{Syn})
contributions -- also known as `information atoms' \cite{Williams2010}.
However, just as the great strength of PID is its capacity to account for
multiple sources of information, its main limitation is that it is restricted
to considering only a single (potentially multivariate) target. Therefore, a
direct application of PID to the TDMI would have to consider the past states of
the system's elements $X_t^1$ and $X_t^2$ as sources and the joint future state
of the system $\bX_{t+1}$ as target. Specifically, focusing on how information
flows from past to future, this account decomposes the information provided by
past states $X_t^1$ and $X_t^2$ about the joint future state $\bX_{t+1}$, as
\begin{align*}
\mathrm{TDMI} = \texttt{Red}(X_t^1,X_t^2;\bX_{t+1} ) 
+ \texttt{Un}(X_t^1;\bX_{t+1}|X_t^2) + \texttt{Un}(X_t^2;\bX_{t+1}|X_t^1) 
+ \texttt{Syn}(X_t^1,X_t^2;\bX_{t+1}) .
\end{align*}
Here, the first term corresponds to the redundant information provided by both
$X_t^1$ and $X_t^2$ about the joint future state of the system $\bX_{t+1}$; the
second and third terms refer to the unique information that only the past state
$X_t^1$ provides about the joint future state $\bX_{t+1}$ (and likewise for
$X_t^2$); and finally, the last term accounts for the synergistic information
that the two elements' past states provide about the system's joint future,
only when they are considered together. Unfortunately, this approach neglects
the fact that the parts of the system are distinct not only in the past, but
also in the future -- in other words, it can tell where the information is
coming from, but not where it is going to. One naive solution would be to
consider the time-reverse of the equation above, with both future states as
sources and the joint past state as target (what we refer to as the `backwards'
PID). However, this leaves unsolved the underlying problem that PID cannot
provide a unified decomposition of information across multiple sources and
multiple targets simultaneously.

In order to obtain an encompassing description of the system's dynamics, one
must extend the PID approach to enable multi-target information decomposition.
To address this issue, our strategy is to take a temporal perspective on PID
itself, focusing on how the information encoded by the PID atoms may evolve
over time. For instance, information that was uniquely encoded by one element
of the system in the past may become redundantly encoded by two in the future,
or synergistic information may subsequently become uniquely encoded by one of
the elements -- and so on. This intuition suggests that when decomposing
information flow between past and future in a system of two elements there are
not four, but rather 16 distinct information atoms: each corresponding to a
pair of the original four PID atoms evolving from past to future. Thus, we
denote each \phiid atom as a pair of PID atoms: e.g. the information that was
carried redundantly in the past and becomes synergistic in the future
corresponds to \rts; and the synergistic information in the past that becomes
unique to $X^1_{t+1}$ in the future corresponds to \stx; and so on.

\begin{figure}[t!]
  \centering
  \includegraphics{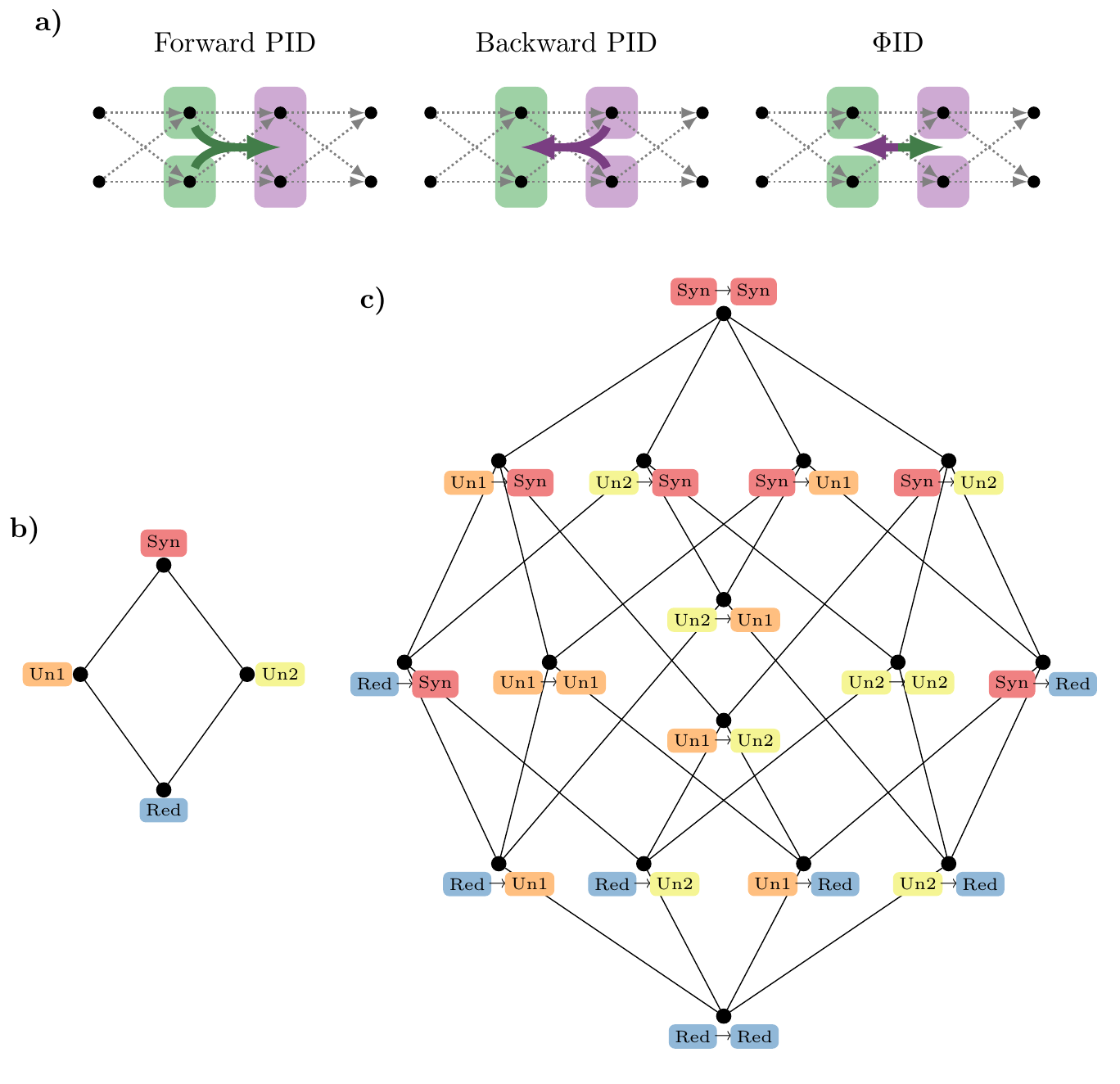}

\caption{\textbf{Lattice representation of information atoms from PID to $\Phi$ID}. 
\textbf{a)} System of two elements co-evolving and
interacting over time, decomposed either according to a \emph{forward PID}
(left) or a \emph{backward PID} (middle).  Integrated Information
Decomposition (\phiid; right) unifies and extends both PIDs, providing an
encompassing framework of information dynamics in complex systems.
 \textbf{b)} Redundant (\texttt{Red}), unique (\texttt{Un}) and synergistic (\texttt{Syn}) atoms in the bivariate PID
lattice. \textbf{c)} \phiid lattice for a system of two time series, where each \phiid atom
corresponds to a pair of two PID atoms that indicate how information evolves from past to future.}

  \label{fig:lattices}
\end{figure}

Like the original PID atoms, the \phiid atoms are structured in a
\emph{lattice}, depicted in Fig.~\ref{fig:lattices} (for a formal derivation
see \textit{Methods}). As with PID, several extra ingredients need to be
specified to compute the numerical value of these atoms: in \phiid, it can be
shown that standard Shannon theory and PID together specify a system of 15
equations for the 16 \phiid atoms, yielding an underdetermined system. To
compute the atoms one must provide one extra constraint, which can be done by
introducing a \emph{double-redundancy} function -- a multi-target extension of
PID's redundancy function.\footnote{For PID, it is often the redundancy
function that provides the needed constraint to allow computation of numerical
values for the PID atoms.} Note that the $\Phi$ID framework does not impose a
particular functional form for the double-redundancy function, and hence
different functional forms can be explored. A formal development of these
ideas, and its extension to systems of more than two elements, is provided in
the Methods.

\subsubsection*{A new taxonomy for information dynamics in complex systems}

Based on $\Phi$ID, we propose an extended taxonomy of information dynamics
according to six disjoint and qualitatively distinct phenomena
(Fig.~\ref{fig:molecules}):

\begin{figure}[t!]
  \centering
  \includegraphics{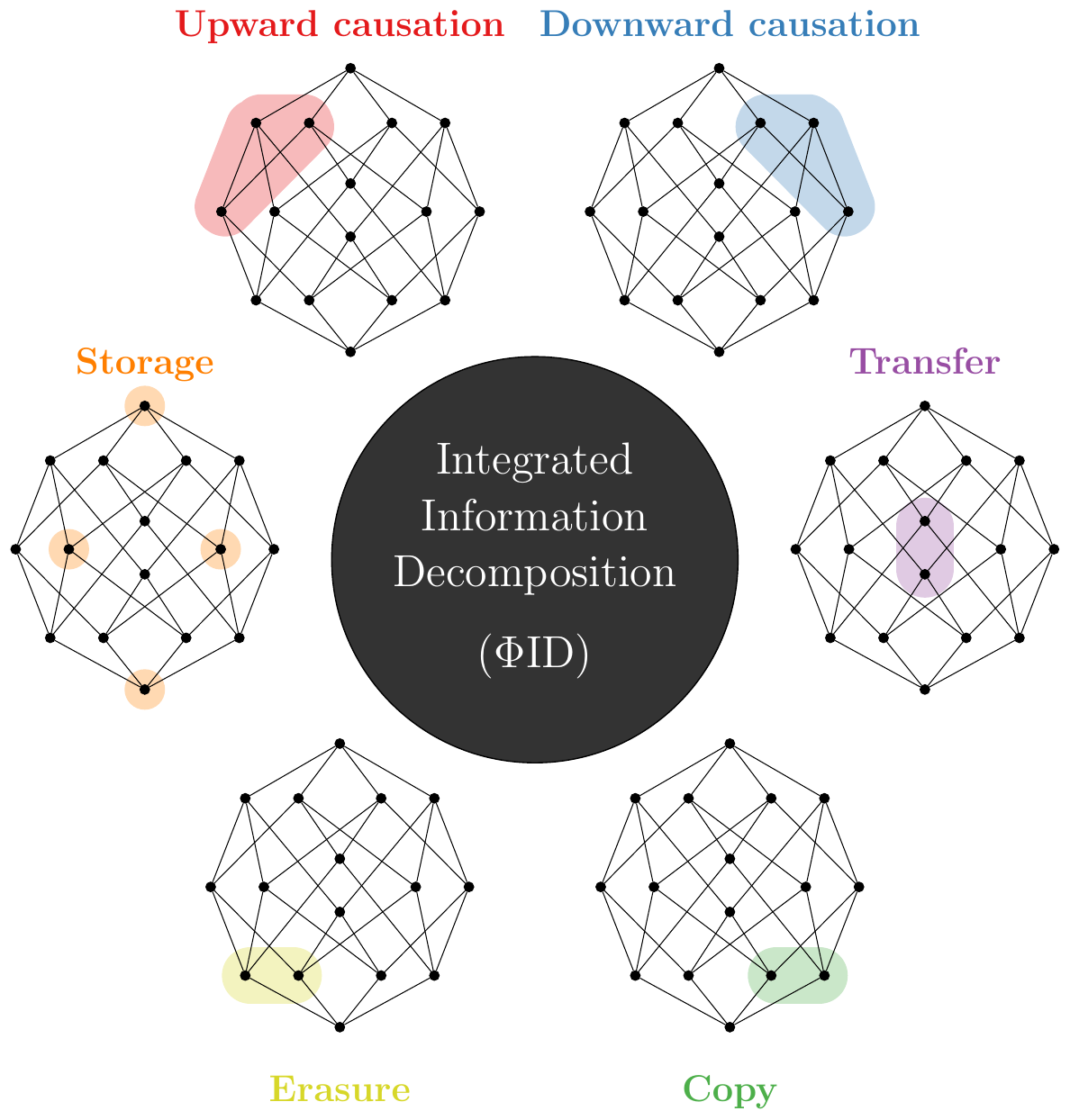}
  \caption{\textbf{Taxonomy of information dynamics in complex systems}. Six qualitatively different modes of information dynamics, represented in terms of their constituent atoms in the \phiid lattice.}
  \label{fig:molecules}
\end{figure}

\begin{description}

\item[Storage]: Information that remains in the same element or set of elements (even if it
includes collective effect). Comprises \rtr, \xtx, \yty, and \sts.

\item[Copy]: Information that becomes duplicated. Comprises \xtr, and \ytr.

\item[Transfer] Information that moves between elements. Comprises \xty and \ytx.

\item[Erasure] Duplicated information that is pruned. Comprises 
\rtx and \rty.

\item[Downward causation] Collective properties that define individual
futures. Comprises \stx, \sty, and \str.

\item[Upward causation] Collective properties that are defined by individuals.
Comprises \xts, \yts, and \rts.

\end{description}

While the downward causation mode has been discussed in the
past~\cite{james2016information}, upward causation and synergistic storage
(\sts) have, to our knowledge, not been reported in the literature.
Importantly, traditional methods of causal discovery cannot capture the type of
modes that involves interactions between multiple target variables. In effect,
while methods such as PID or multivariate Granger causality can effectively
deal with multivariate target variables, they cannot untangle how each
component of the target may be differently affected, and --- more importantly
--- how sources may affect the target as a whole, without (or in addition to)
affecting its parts. Overall, this new taxonomy leads to less ambiguous and
fully quantifiable descriptions of information dynamics in complex systems, in
addition to grounding abstract concepts such as upward and downward
causation,\footnote{The relation between $\Phi$ID and causal
emergence~\cite{seth2010measuring} can be found in a separate
publication~\cite{Rosas2020}.} and notions such as integrated information -- as
we discuss below.

\subsubsection*{A simple example of information decomposition with $\Phi$ID}

As a first example of the kind of insight that Integrated Information
Decomposition can provide, let us focus on the decomposition of a variable's
so-called `active information storage' (AIS)~\cite{Lizier2010}, which is
defined as the TDMI of an individual part of the system (i.e. the mutual
information between the present of one variable, $X^1_{t}$, and its own future,
$X^1_{t+1}$). To decompose AIS, consider that in PID the mutual information of
a single source variable with the target is decomposed as the sum of redundancy
(which is information that each source has about the target) and that source's
unique information. Similarly, in \phiid AIS is decomposed in terms of
redundancy and unique information, but now taking into account \emph{both} past
and future:
\begin{align}
\begin{split}
\mathrm{AIS}(X^1) = I(X^1_{t}; X^1_{t+1}) = \rtr \,+\, \rtx \,+\, \xtr \,+\, \xtx ~ .
\label{eq:ais}%
\end{split}
\end{align}
Here, \rtr corresponds to redundant information in the past of both parts that
is present in the future of both parts; \rtx is the redundant information in
the past that is eliminated from the second element and hence is only conserved
in $X^1_{t+1}$; and similarly for the remaining atoms.

Even with this simple example, \phiid already yields new insights into the
system's information dynamics: note that, although $X^2_{t},X^2_{t+1}$ are not
in this mutual information, $I(X^1_{t};X^1_{t+1})$ shares the \rtr term with
$I(X^2_{t};X^2_{t+1})$ by virtue of them being considered part of the same
multivariate stochastic process. Therefore, if one uses simple mutual
information as a measure of storage one may include information that is not
stored exclusively in a given variable, resulting in a `double-counting' of the
\rtr atom which can lead to paradoxical conclusions -- such as the sum of
individual information storages being greater than the total information flow
(TDMI).

\begin{figure}[ht]
    \centering
    \includegraphics{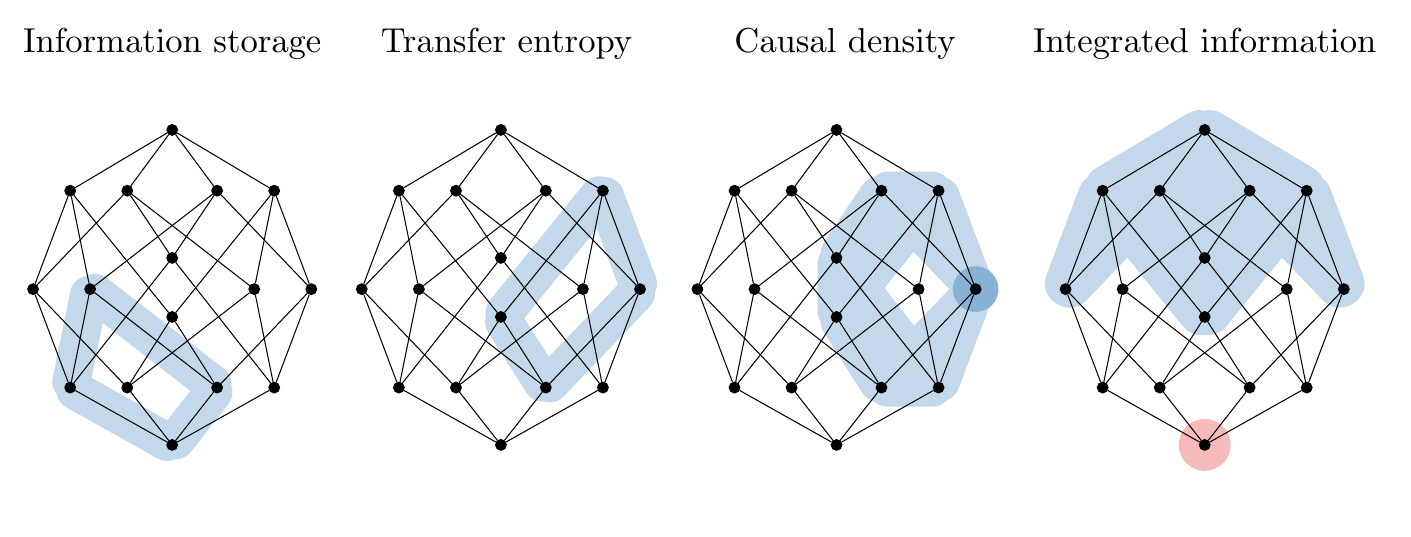}
    \caption{\textbf{Common information-theoretic measures decomposed into integrated information atoms}. Constituent \phiid atoms of active information storage, transfer entropy (from $X_t^1$ to $X_{t+1}^2$), causal density (sum of transfer entropies), and whole-minus-sum integrated information, highlighted in blue. Dark blue indicates double-counting in causal density, and red indicates a negative contribution of redundancy to integrated information (see text for details).}
    \label{fig:phimolecules}
\end{figure}

More generally, \phiid can be used decompose many other quantities of interest for
complex systems analysis (Fig.~\ref{fig:phimolecules}), and their decompositions can
help us both to understand existing measures and design new ones.
In the following sections we apply this line of reasoning and the \phiid framework
to two prominent scenarios in complex systems analysis: the assessment of causal
interactions between system components, and the quantification of system-wide integrated
information.

\subsection*{Theoretical implications}

\subsubsection*{Different types of integration}

Measures of integrated information, usually denoted by $\Phi$, aim to quantify
the degree to which a temporal evolution of a dynamical system depends on the
interdependencies between its parts~\cite{Balduzzi2008},\footnote{The parts
being chosen so as to have the weakest overall informational link between
them.} see \cite{Mediano2019} for a review. Integrated information measures
have been applied widely, most notably in the neuroscience of consciousness,
but also to studies of diverse complex
systems~\cite{Mediano2016,Aguilera2019,mediano2021integrated}. In this section
we investigate the concept of integrated information through the lens of
\phiid.

\begin{figure}[h]
  \centering
  \includegraphics{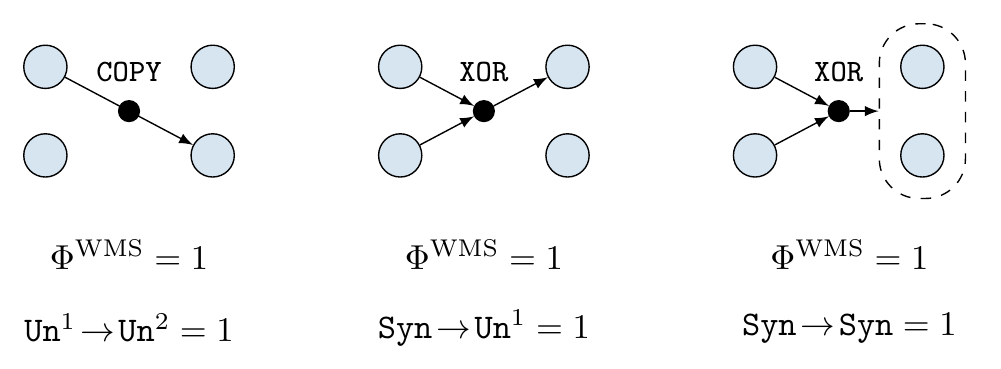}
  \caption{\textbf{Example systems of logic gates}. While these three systems have the same integrated
  information (measured with \wmsphi), their information dynamics are radically 
different. The idiosyncrasy of each type of dynamic is captured by the \phiid formalism,
which shows that in each system there is only one non-zero atom, different for each system.}
  \label{fig:examples}
\end{figure}

The key insight that $\Phi$ID delivers is that there are multiple qualitatively
different ways in which a multivariate dynamical process can integrate
information -- through different combinations of redundant, unique, and
synergistic effects. To illustrate this, let's focus on the so-called
"whole-minus-sum" empirical integrated information metric~\cite{Barrett2011},
which for a simple 2 component system is calculated as\footnote{There is only
one possible partitioning of a 2 component system, so here we don't need to
search for the minimum information partition.}
\begin{align}
  \Phi^{\mathrm{WMS}} = I(\bX_t ; \bX_{t+1}) - \sum_{i} I(X_t^i; X_{t+1}^i) ~ .
  \label{eq:wmsphi}%
\end{align}
which reflects a balance between the information contained within the whole
system ($I(\bX_t ; \bX_{t+1})$) and the information contained within the parts
($I(X_t^1; X_{t+1}^1)$ and $I(X_t^2; X_{t+1}^2)$). We apply this metric to the
following elementary examples of 2 binary variables (Fig.~\ref{fig:examples}):

\begin{itemize}

  \item A \textbf{copy transfer} system, in which $X_t^1,X_t^2,X_{t+1}^1$ are
  i.i.d. fair coin flips, and $X_{t+1}^2 = X_t^1$ (i.e. the information of $X_t^1$
  is copied to $X_{t+1}^2$).

  \item The \textbf{downward XOR}, in
  which $X_t^1,X_t^2,X_{t+1}^2$ are i.i.d. fair coin flips, and $X_{t+1}^1 \equiv X_t^1 + X_t^2 ~\mathrm{(mod~2)}$.

  \item The \textbf{parity-preserving random} (PPR), in
  which $X_t^1,X_t^2$ are i.i.d. fair coin flips, and $X_{t+1}^1 + X_{t+1}^2
  \equiv X_t^1 + X_t^2~\mathrm{(mod~2)}$ (i.e. $\bm X_{t+1}$ is a random string of the same parity as $\bm X_t$).

\end{itemize}

A direct calculation shows that these three systems are `equally integrated':
$\wmsphi=1$ for all of them, which implies that the degree to which the
dynamics of the whole cannot be perfectly predicted from the parts alone is
equivalent~\cite{Mediano2019,Barrett2011}. However, a more nuanced analysis
using $\Phi$ID reveals that these systems integrate information in
qualitatively different ways. In effect, the integration in the copy system is
entirely due to transfer dynamics (\xty); the downward \texttt{XOR} integrates
information by transforming synergistic into unique information (\stx); and PPR
due to information synergistic in the past and future (\sts). All the other
\phiid atoms in each of these systems are zero (proofs in the Appendix).

\subsubsection*{Measures of integrated information capture multiple \phiid atoms }

Within the IIT literature, researchers have proposed multiple measures aimed at
quantifying to what extent a system is integrated as a whole, in terms of its
parts influencing each other's evolution over time~\cite{Tegmark2016}. These
measures, though superficially similar, are known to behave inconsistently, for
reasons that are not always clear~\cite{Mediano2019}. Here we use $\Phi$ID to
dissect and compare three existing measures of integrated information
($\wmsphi$, $\psi$, $\Phi_G$) and causal density (CD), bringing to light their
similarities and differences.\footnote{We provide definitions of each measure
in the Supplementary Material -- for details see Section 2.2 of
Ref.~\cite{Mediano2019} and the original
references~\cite{Balduzzi2008,Griffith2014,Oizumi2016}.}

As a systematic exploration, one can determine which measures are sensitive to
which modes of information dynamics by calculating whether each measure is
zero, positive, or negative for a system consisting of only one particular
$\Phi$ID atom (Table~\ref{tab:comparison}; proofs in the Appendix). Strikingly,
each proposed measure of integration captures a distinct combination of
$\Phi$ID atoms: although generally most of them capture synergistic effects and
avoid (or penalise) redundant effects, they differ substantially.

The conclusion of this analysis is that these measures are not simply different
approximations of a single concept of integration, but rather they are
capturing intrinsically different aspects of the system's information dynamics.
While aggregate measures like these can be empirically useful, one should keep
in mind that they are measuring combinations of different effects within the
system's information dynamics. Echoing the conclusions of
Ref.~\cite{Mediano2019}: these measures behave differently not only in
practice, but also \emph{in principle}.

\vspace{-5pt}
\begin{table}[h]
  \caption{\textbf{Sensitivity of integrated information measures to $\Phi$ID atoms}. For each measure, entries indicate whether the value is positive (+), negative (-) or 0 in a system in which the given $\Phi$ID atom is the only non-zero atom. Atom colour code taken from Fig.~\ref{fig:lattices}.}
  \label{tab:comparison}
  \centering
  \colorlet{RedCol}{BrewerBlue!55}
  \colorlet{SynCol}{BrewerRed!55}
  \colorlet{Un1Col}{BrewerOrange!50}
  \colorlet{Un2Col}{BrewerYellow!50}
  \newcommand\pad{\vphantom{Sy\textsuperscript{i}}}
  \renewcommand{\arraystretch}{1.6}
  \setlength\tabcolsep{5pt}
  \def\l{0.55cm}
  \def\m{0.1em}
  \vspace{5pt}
  \begin{tabular}{l | c c c c}
    \multicolumn{1}{c}{\bfseries $\Phi$ID atoms} & \multicolumn{4}{c}{\textbf{Measures}} \\
    ~                                & $\wmsphi$ & $\Phi_G$ & $\psi$ & \textrm{CD} \\
    \toprule                                                         
    \lfbox[width=\l, padding-break-top=\m, padding-break-bottom=\m, rounded, background-color=SynCol, border-width=0pt]{\pad\texttt{Syn}}$\rightarrow$\lfbox[width=\l, padding-break-top=\m, padding-break-bottom=\m, rounded, background-color=SynCol, border-width=0pt]{\pad\texttt{Syn}}              &     +     &    0     &    +   &      0      \\
    \lfbox[width=\l, padding-break-top=\m, padding-break-bottom=\m, rounded, background-color=SynCol, border-width=0pt]{\pad\texttt{Syn}}$\rightarrow$\lfbox[width=\l, padding-break-top=\m, padding-break-bottom=\m, rounded, background-color=Un1Col, border-width=0pt]{\pad\texttt{Un}\textsuperscript{i}}              &     +     &    +     &    +   &      +      \\
    \lfbox[width=\l, padding-break-top=\m, padding-break-bottom=\m, rounded, background-color=SynCol, border-width=0pt]{\pad\texttt{Syn}}$\rightarrow$\lfbox[width=\l, padding-break-top=\m, padding-break-bottom=\m, rounded, background-color=RedCol, border-width=0pt]{\pad\texttt{Red}}              &     +     &    +     &    +   &      +      \\
    \lfbox[width=\l, padding-break-top=\m, padding-break-bottom=\m, rounded, background-color=Un1Col, border-width=0pt]{\pad\texttt{Un}\textsuperscript{i}}$\rightarrow$\lfbox[width=\l, padding-break-top=\m, padding-break-bottom=\m, rounded, background-color=SynCol, border-width=0pt]{\pad\texttt{Syn}}              &     +     &    0     &    0   &      0      \\
    \lfbox[width=\l, padding-break-top=\m, padding-break-bottom=\m, rounded, background-color=RedCol, border-width=0pt]{\pad\texttt{Red}}$\rightarrow$\lfbox[width=\l, padding-break-top=\m, padding-break-bottom=\m, rounded, background-color=SynCol, border-width=0pt]{\pad\texttt{Syn}}              &     +     &    0     &    0   &      0      \\
    \lfbox[width=\l, padding-break-top=\m, padding-break-bottom=\m, rounded, background-color=Un1Col, border-width=0pt]{\pad\texttt{Un}\textsuperscript{i}}$\rightarrow$\lfbox[width=\l, padding-break-top=\m, padding-break-bottom=\m, rounded, background-color=Un1Col, border-width=0pt]{\pad\texttt{Un}\textsuperscript{i}}              &     0     &    0     &    0   &      0      \\
    \lfbox[width=\l, padding-break-top=\m, padding-break-bottom=\m, rounded, background-color=Un1Col, border-width=0pt]{\pad\texttt{Un}\textsuperscript{i}}$\rightarrow$\lfbox[width=\l, padding-break-top=\m, padding-break-bottom=\m, rounded, background-color=Un2Col, border-width=0pt]{\pad\texttt{Un}\textsuperscript{j}}              &     +     &    +     &    0   &      +      \\
    \lfbox[width=\l, padding-break-top=\m, padding-break-bottom=\m, rounded, background-color=Un1Col, border-width=0pt]{\pad\texttt{Un}\textsuperscript{i}}$\rightarrow$\lfbox[width=\l, padding-break-top=\m, padding-break-bottom=\m, rounded, background-color=RedCol, border-width=0pt]{\pad\texttt{Red}}              &     0     &    +     &    0   &      +      \\
    \lfbox[width=\l, padding-break-top=\m, padding-break-bottom=\m, rounded, background-color=RedCol, border-width=0pt]{\pad\texttt{Red}}$\rightarrow$\lfbox[width=\l, padding-break-top=\m, padding-break-bottom=\m, rounded, background-color=Un1Col, border-width=0pt]{\pad\texttt{Un}\textsuperscript{i}}              &     0     &    0     &    0   &      0      \\
    \lfbox[width=\l, padding-break-top=\m, padding-break-bottom=\m, rounded, background-color=RedCol, border-width=0pt]{\pad\texttt{Red}}$\rightarrow$\lfbox[width=\l, padding-break-top=\m, padding-break-bottom=\m, rounded, background-color=RedCol, border-width=0pt]{\pad\texttt{Red}}              &    $-$    &    0     &    0   &      0      \\
  \end{tabular}
\end{table}

\subsubsection*{A $\Phi$ID account of information transfer}

Most methods of statistical causal discovery use the conditional mutual
information\footnote{Or linear variants of it, to which our conclusions also
apply.} as their main building block. Here we illustrate how one of such
approaches, transfer entropy (TE)~\cite{bressler2011wiener}, can be decomposed
in terms of $\Phi$ID, showing that it conflates qualitatively distinct effects
in non-straightforward ways, and fails to capture high-order modes of
information flow. The TE from the system's first to its second element, defined
as $\text{TE}(1\rightarrow 2) := I(X^1_t; X^2_{t+1}|X^2_t)$, can be decomposed
via $\Phi$ID as
\begin{align}
\begin{split}
  \text{TE}(1\rightarrow 2) = \str \,+\, \sty \,+\, \xtr \,+\, \xty ~ .
\label{eq:te}%
\end{split}
\end{align}
Note that, of these, \xty is the only `genuine' transfer term -- all others
correspond to redundant or synergistic effects involving both variables in past
or future, which do not imply any kind of transfer phenomena. This complicates
the interpretation of TE as a fully general measure of information
transfer.\footnote{Similar concerns about TE have been raised in
Ref.~\cite{james2016information}.} In contrast, $\Phi$ID can isolate the part
of the transfer entropy that corresponds to information transfer through the
\xty term.

Additionally, Eq.~\eqref{eq:te} implies that the atom \str is accounted for in
both $\text{TE}(1\rightarrow 2)$ and $\text{TE}(2\rightarrow 1)$. This has an
important consequence: if one quantifies the total causal influence within the
system via adding up both TEs (a quantity known as \textit{unnormalised causal
density} (uCD)~\cite{Mediano2019}), this may overestimate the effective
interdependencies by double-counting this atom (Fig.~\ref{fig:phimolecules}).
In fact, the double-counting of \str implies that $\text{uCD}$ can potentially
be even larger than TDMI. While this overestimation of uCD has been noted
before~\cite{Oizumi2016}, $\Phi$ID not only reveals the precise reason for this
overestimation, but also provides a practical solution: one can correct uCD by
substracting the double-counted $\Phi$ID atom. This problem -- and its solution
-- may have important consequences in fields such as computational
neuroscience, where the total TE of a brain region is a popular metric of its
relevance for the brain's hierarchical
organization~\cite{deco2021revisiting,luppi2020synergisticConsciousness}.

The decomposition of uCD via $\Phi$ID reveals another important limitation of
traditional causal discovery methods: they do not account for modes of
information flow that involve synergy in the targets. Therefore, while these
methods account for possible interactions of source variables, they neglect
interactions in the targets -- which are never considered jointly. As an example,
take the parity-preserving system in Fig.~\ref{fig:examples}: this system has zero
CD and zero AIS, yet it clearly has dynamical structure (as picked up by \wmsphi).
This is an example of information being carried purely in high-order effects, in a
way that common measures like AIS and TE are unable to capture.
More generally, we expect this to be particularly prevalent in systems with
distinct micro- and macro-scale behaviour~\cite{seth2010measuring,Rosas2020}.

\subsection*{Numerical examples}
\label{sec:numerical}

In this section we showcase three applications of \phiid to simulated and real
data, illustrating the capabilities of \phiid to yield new insights and solve
practical problems. As stated above, the numerical calculation of \phiid atoms
depends on a choice of double-redundancy function -- which, as in the case of
PID, gives room to a range of options (see e.g.
Refs~\cite{Williams2010,Ince2017,Barrett2015,finn2018pointwise}). In all
examples below we use a multi-target extension of the Common Change in
Surprisal (CCS) measure by Ince~\cite{Ince2017}; furthermore, we show that our
all the results replicate with a multi-target extension of Barrett's Minimum
Mutual Information (MMI) measure~\cite{Barrett2015} (see Supplementary
Material).

\subsubsection*{Why whole-minus-sum $\Phi$ can be negative}
\label{sec:wmsphi}

\phiid can be further leveraged to explain certain behaviours of integrated
information and dynamical complexity measures. In particular, \wmsphi can
sometimes take negative values, which could suggest a counter-intuitive notion
of a `negatively integrated' system. In fact, this behaviour has been used as
an argument to discard \wmsphi as a suitable measure of integrated
information~\cite{Griffith2014,Oizumi2016}. \phiid can provide an explanation
of this otherwise paradoxical behaviour, and furnishes a simple solution.

By applying \phiid to the definition of \wmsphi in Eq.~\eqref{eq:wmsphi}, one
finds that \wmsphi accounts for all the synergies in the system, the unique
information transferred between parts of the system and, importantly, the
negative of the \rtr atom (Fig.~\ref{fig:phimolecules}; see Supplementary
Material for details). The presence of this negative double-redundancy term
shows that \wmsphi can be negative in highly redundant systems, in which \rtr
is larger than all other atoms that constitute \wmsphi. This is akin to
Williams and Beer's~\cite{Williams2010} explanation of the negativity of the
interaction information~\cite{mcgill1954multivariate}, applied to multivariate
processes. Based on this insight, one can formulate a `revised' measure of
integrated information, \phiR, by adding back the double-redundancy, which
includes only synergistic and unique transfer terms.

We computed \phiR numerically for a simple two-node autoregressive (AR) system,
mimicking the setting in Ref.~\cite{Mediano2019}. The system consists of two
continuous variables with dynamics such that $\bm X_{t+1} \sim \mathcal{N}(A
\bm X_t; \Sigma)$, with $A$ being a $2 \times 2$ matrix with all entries set to
0.4, and $\Sigma$ a noise (or \emph{innovations}) covariance matrix with 1's
along the diagonal and a given noise correlation ($c$ in the notation of
Ref.~\cite{Mediano2019}) in the off-diagonal entries. We calculated
\wmsphi and \phiR with respect to the system's stationary
distribution, which can be shown to be a multivariate Gaussian. Plots of both
quantities are shown in Fig.~\ref{fig:wms_ar}, and details of the computation
can be found in the Appendix.

\begin{figure}[ht]
  \centering
  \includegraphics{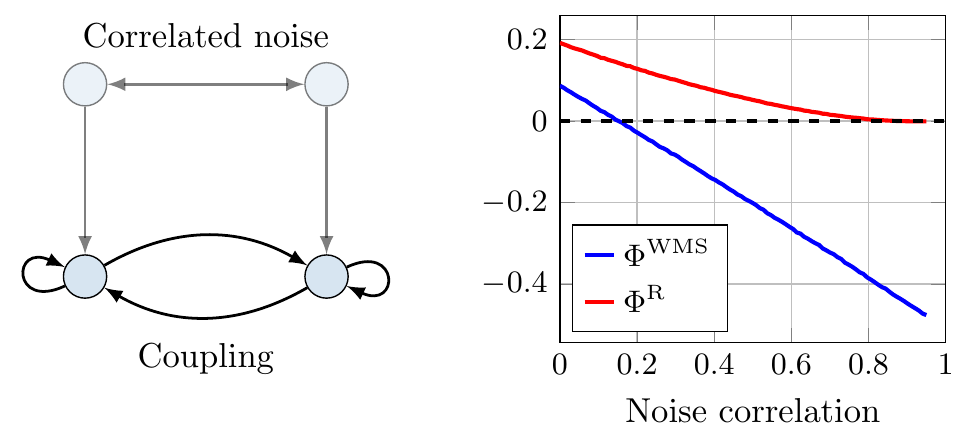}

  \caption{\textbf{Standard and revised \wmsphi in a two-component noisy autoregressive system}. As the noise injected to both components becomes more correlated, \wmsphi drops below zero while \phiR remains positive.}

  \label{fig:wms_ar}
\end{figure}

As expected, \wmsphi drops below zero as synergy decreases and redundancy
increases with noise correlation. However, after adding back the
double-redundancy term, the revised version, \phiR, tends to 0 for high noise
correlation, which is more consistent with some of the other measures
highlighted in Ref.~\cite{Mediano2019}, e.g.~CD and $\Phi^*$.

\subsubsection*{Information decomposition in simulated whole-brain activity}

We next analysed simulated whole-brain activity via the well-known Dynamic Mean
Field (DMF) model introduced by Deco \emph{et al.}~\cite{Deco2014}. This model
represents cortical regions as macroscopic neural fields, whose local dynamics
are described by a set of coupled differential equations. The DMF model
incorporates realistic aspects of neurophysiology such as synaptic dynamics and
membrane potential~\cite{deco2012ongoing,deco2013resting,hansen2015functional},
and is informed by a network of anatomical connections obtained e.g. from
diffusion tensor imaging (DTI) - while having the advantage of being free from
physiological noise confounds. An additional biophysical haemodynamic
model~\cite{friston2000nonlinear} enables the firing rates generated by the DMF
model to be transformed into BOLD signals similar to resting-state fMRI data
from humans, and that have been subject of study in applications of the DMF to
model the neural effects of sleep~\cite{ipina2020modeling},
anaesthesia~\cite{luppi2021paths}, and psychedelic drugs~\cite{Herzog2020}.

\begin{figure}[t]
  \begin{adjustwidth}{-1.25in}{0in}
    \includegraphics{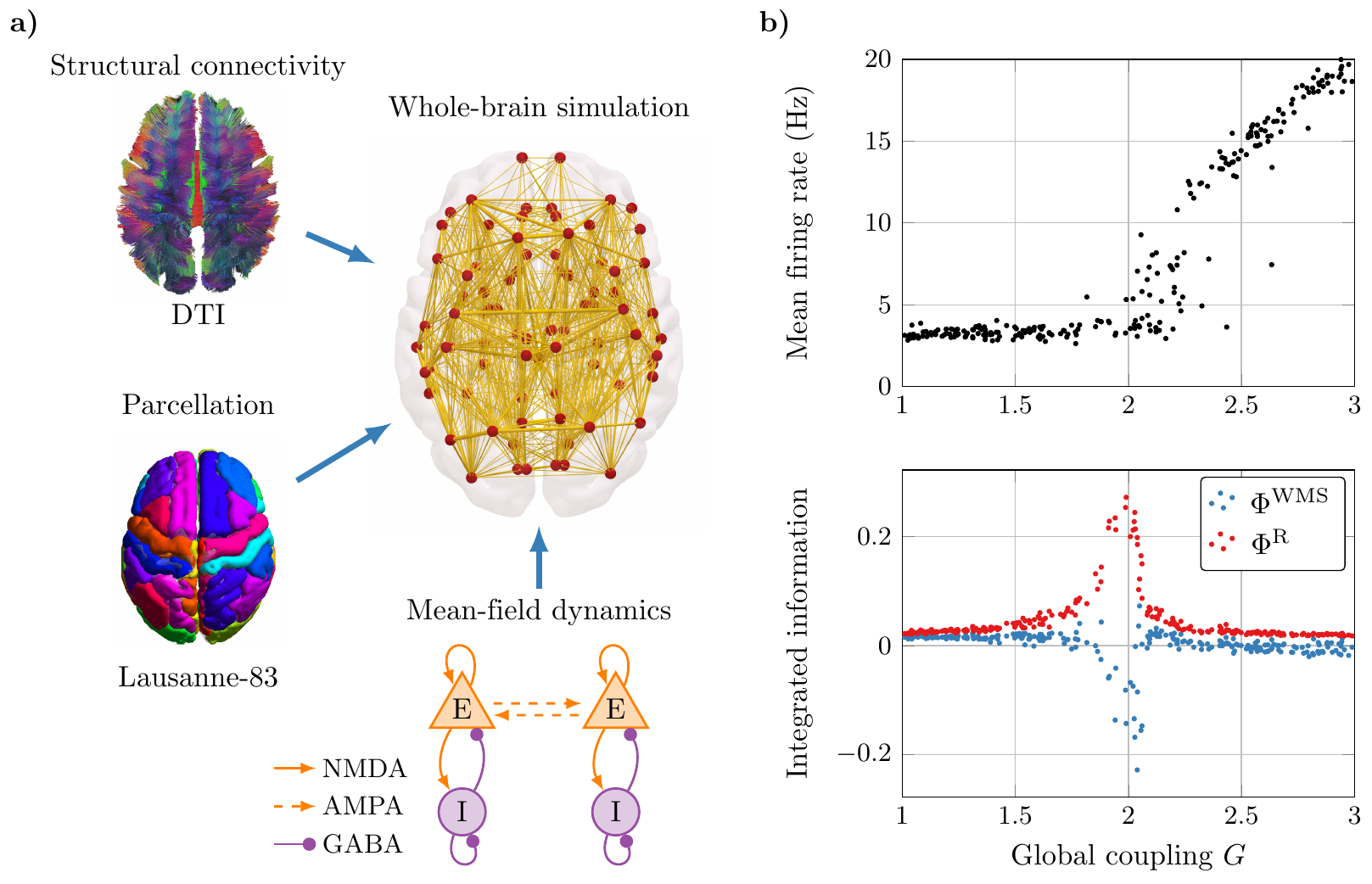}
  \end{adjustwidth}

\caption{\textbf{Measures of integrated information in a whole-brain
computational model}. \textbf{a)} Schematic diagram of the Dynamic Mean-Field
(DMF) model~\cite{Deco2014,luppi2021paths}, which combines a DTI-based connectome and a
whole-brain parcellation to simulate realistic BOLD signals. \textbf{b)} As the
global coupling parameter $G$ is increased, the mean firing rate exhibits a sharp increase at approximately $G = 2$. Importantly, \wmsphi shows a downward peak,
suggesting a conceptually problematic negative value of integration -- while the revised measure
\phiR shows a strong positive peak.}

  \label{fig:dmf}
\end{figure}

We simulate the DMF equations using a DTI-based connectome obtained from the
public Human Connectome Project data~\cite{Glasser2013} using the same model
settings as Herzog \emph{et al.}~\cite{Herzog2020}, and compute
$\Phi^{\text{WMS}}$ and $\Phi^{\text{R}}$ for all pairs of brain
regions.\footnote{The choice of analysing pairs of regions is only for
convenience, as the theory is defined to systems of any size.} These values
were calculated for varying values of a global coupling parameter (denoted by
$G$) and the resulting average values (over all pairs of brain regions) were
then analysed. The details of the model, the simulation procedure, and the
computation of integrated information measures can be found in the Appendix.

As shown in Fig.~\ref{fig:dmf}, for values of $G$ close to 2 the mean firing
rate of the model increases sharply, reminiscent of a phase transition. At this
point \wmsphi shows a marked decrease, suggesting that the system is least
integrated in the transition region. In fact, the value of \wmsphi is often
less than zero, which would correspond to the conceptually problematic notion
of a system that is 'negatively integrated.' Crucially, however, when the
double-counting of redundancy is corrected and \phiR is used instead, a
completely different picture appears: integration (understood as synergy plus
transfer) is always positive, and it strongly \emph{increases} and peaks in the
transition region. This result aligns well with prior
literature~\cite{Deco2018,Herzog2020} showing that the point $G=2$ corresponds
to the model's optimal fit to data from awake subjects, and that a high level
of integration is required for the normal operation of the brain in healthy,
conscious individuals~\cite{Tononi1994}.

The strong discrepancy between \wmsphi and \phiR confirms that the concerns
regarding the conflation of multiple information effects highlighted throughout
this article are not a mere theoretical issue, but can trigger misleading
interpretations in the analysis of neuroscientific data. Hence, this example
illustrates the capability of \phiid to disambiguate between qualitatively
different dynamical phenomena in time series data.

\subsubsection*{$\Phi$ID sheds new light on empirical results}

To provide an empirical demonstration of the capabilities of $\Phi$ID, we used
it to study the dynamical relationship that exists between heart rate and
respiration in healthy human subjects. This choice was motivated by the
well-known influence of respiration on heart rate, which can be captured in
terms of transfer entropy between respiratory volume and heart rate time
series~\cite{nemati2013respiration,pini2019lagged,de2020transfer}. Therefore,
we sought to investigate how the $\Phi$ID framework could be used to decompose
this effect into its constituent informational elements.

\begin{figure}[ht]
  \centering
  \includegraphics{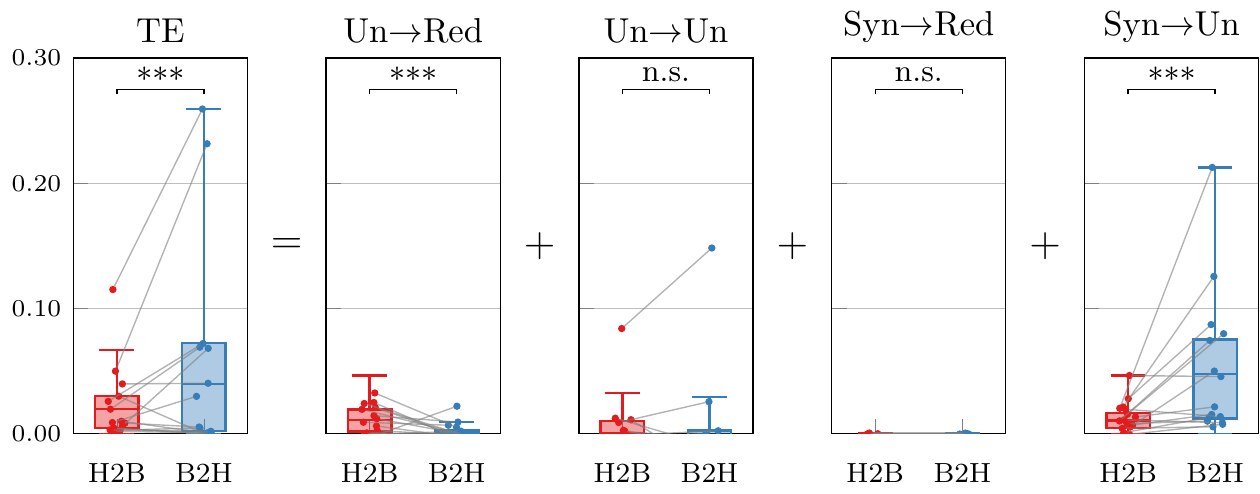}
  \caption{\textbf{Decomposition of information transfer between heart rate and respiratory volume}. Transfer entropy decomposed into its four constituent \phiid atoms, calculated in both directions (heart to breath [H2B] and breath to heart [B2H]).}
  \label{fig:fantasia}
\end{figure}

For this purpose, we analysed the Fantasia database~\cite{iyengar1996age}, an
openly available dataset that contains data from 40 healthy subjects while
watching the Disney movie ‘Fantasia.’ Following the preprocessing pipeline
outlined in Ref.~\cite{nemati2013respiration}, we extracted synchronised time
series for inter-beat intervals from the ECG timeseries, and detrended the
respiratory volume. Using the resulting data, we calculated both the transfer
entropy from heart to breath and from breath to heart, and then proceeded to
decompose these quantities in terms of their $\Phi$ID constituents
(Fig.~\ref{fig:fantasia}).

As expected based on previous work, the transfer entropy from breath to heart
was significantly higher than from heart to breath. Crucially, our analysis
revealed that this effect is driven by two distinct $\Phi$ID atoms, out of the
four that comprise transfer entropy. The TE result is dominated by the
$\texttt{Syn}\!\rightarrow\!\texttt{Un}$ atom, while the transfer atom itself
($\texttt{Un}\!\rightarrow\!\texttt{Un}$) shows no significant differences.
Importantly, however, $\texttt{Un}\!\rightarrow\!\texttt{Red}$ shows a
significant effect in the \emph{opposite} direction to the main TE result. The
standard TE analysis is unable to resolve these modes of information dynamics,
and thus misses the heart's unique contribution to the heart-breath joint
dynamics.

In summary, $\Phi$ID shows that the effect seen in the transfer entropy towards
the heart is not transfer but synergistic, and that there is a smaller unique
effect originating at the heart that is overshadowed by the
former.\footnote{The study of the physiological implications of these findings
will be investigated in a separate publication.} These results illustrate the
type of advantages that $\Phi$ID can bring beyond standard transfer entropy and
Granger causality analyses.

\section*{Discussion}

This paper introduces $\Phi$ID as a formal framework to study high-order
interactions in the dynamics of multivariate complex systems. By bringing
together aspects of integrated information theory (IIT) and partial information
decomposition (PID), the $\Phi$ID framework allows us to decompose multivariate
information flow into interpretable, distinct parts. This decomposition brings
two important outcomes. First, it allows us to better understand and refine
existing metrics of information exchange and dynamical complexity. Second, it
enables systematic analyses of previously unexplored modes of information
dynamics, which are not captured by previous analysis methods.

\subsection*{Towards multi-dimensional accounts of dynamical complexity}

$\Phi$ID provides principled tools to inspect existing measures of information
dynamics and overcome some of their shortcomings. In particular, we have shown
that both the widely used transfer entropy and what is referred to as
`integrated information' in the context of IIT are in fact aggregates of
several distinct information effects, typically including transfer and synergy
phenomena. In addition, our analysis shows that different measures of
integrated information actually capture different $\Phi$ID atoms in various
proportions, which provides a formal explanation for the heterogeneity among
existing measures reported in Ref.~\cite{Mediano2019}.

Supporting our theoretical results, we showed in empirical physiological data
that significant differences in transfer entropy can be observed even in the
absence of genuine information transfer phenomena, highlighting the practical
relevance of our framework. Likewise, our analysis based on whole-brain
modelling showed how the conflation of fundamentally distinct information
phenomena in existing measures of integrated information can introduce
substantial confusion in the results and interpretation of neuroscientific
analysis. $\Phi$ID provides tools to identify these problems, and also to fix
them, by tailoring measures that target the specific kinds of information
dynamics one wishes to analyse.

As well as providing a new taxonomy of information dynamics phenomena, $\Phi$ID
establishes that there are fundamental limitations to any purported
all-encompassing scalar measure of dynamical complexity, in line with Feldman
and Crutchfield~\cite{Feldman1998}. The space of possible complex dynamics is,
unsurprisingly, vast and complex, and while scalar measures might still have
great practical value in specific contexts,\footnote{For example, measures that
accurately discriminate between neural configurations corresponding to
conscious and unconscious states in a particular experimental
paradigm~\cite{Casali2013}.} $\Phi$ID clarifies that a general theory of
complex systems (biological or otherwise) cannot be reduced to a single,
one-size-fits-all measure, but rather needs to embrace this richness.

\subsection*{Limitations and future extensions}

$\Phi$ID is a general tool to decompose multivariate mutual information, and
the nature of the resulting decomposition critically depends on how the
underlying joint distribution has been constructed. In particular, note that
the \phiid framework depends only on a joint probability distribution $p(\bX_t,
\bX_{t+1})$, and thus its results can be interpreted as causality in the Pearl
or Granger sense, depending on whether the distribution comes from intervention
or observation, respectively. If the distribution is built on observational
data then the decomposition generally should be understood in the
Granger-causal sense (i.e. as referring to predictive ability). Similarly, if
the conditional distribution $p(\bX_{t+1}|\bX_t)$ is equivalent to a
\texttt{do()} distribution in Pearl's sense~\cite{pearl2018book}, and the
system satisfies the faithfulness and causal Markov conditions, then the
results of $\Phi$ID are to be interpreted in a counterfactual causal sense. In
either case, the formalism developed here applies directly.

Naturally, \phiid inherits some characteristics of PID. In particular, \phiid
specifies the relationships between information atoms, but does not prescribe a
particular functional form for them. To compute numerical values of \phiid
atoms one needs a double-redundancy function, a multi-target extension of PID's
redundancy function. In PID, several distinct redundancy functions have been
proposed, and while they have been shown to agree in various
scenarios~\cite{rosas2020operational}, they may differ in other cases and there
is not yet a consensus on one that is universally preferable~\cite{James2018}.
The formulation of multiple double-redundancy functions, and a thorough
comparison in simulated systems is an important line for future work.

Finally, it is important to remark that the framework presented in this paper
focuses on decomposing the mutual information between two time points. While
this captures all the information carried from past to future in Markovian
systems, it might miss relevant phenomena in systems with non-Markovian
dynamics -- which typically arise in experimental data of scenarios with many
non-observable variables. As an important extension, future work should
investigate the effect of unobserved variables on the proposed decomposition,
which could be done e.g. leveraging Taken’s embedding
theorem~\cite{takens1981detecting,cliff2016information} or other
methods~\cite{wilting2018inferring}.

\section*{Methods}

This section establishes the mathematical bases of our framework. The aim is to
build a decomposition of the TDMI, as defined in
Eq.~\eqref{eq:excess_entropy_markov}, that differentiates the role of each
source and each target (possibly multivariate) variable -- hence accounting for
both cause (forward) and effect (backward) information simultaneously. To do
this, our proposed decomposition brings together the partitions enabled by a
forward PID (where variables at time $t$ and $t+1$ are sources and targets,
respectively) and a backward PID (where the assignment of sources and targets
is reversed), as illustrated in Figure~\subref{fig:lattices}{a}. By doing this,
we overcome PID's limitation of having only one single target variable and
formulate a multi-target information decomposition.

Note that throughout this section we switch from the
\texttt{Red}/\texttt{Un}/\texttt{Syn} notation above to the more standard (and
more general) `curly bracket' notation introduced by Williams and
Beer~\cite{Williams2010}.

\subsection*{Double-redundancy lattice}

Let us start by reviewing the construction of the redundancy lattice that is
employed in PID to formalise our intuitive understanding of
redundancy~\cite{Williams2010}.
This lattice is built over the set $\mathcal{A}$, which for the case of two
time series can be expressed as
\begin{equation}
\mathcal{A} := \{ \{1\}, \{2\}, \{1,2\}, \{ \{1\},\{2\} \} \}, 
\end{equation}
which correspond to all the sets of subsets of $\{1,2\}$ where no element is
contained in another.\footnote{For a case of $N$ variables, then $\mathcal{A}$
is the set of antichains on the lattice $(\mathcal{P}(\{1, ... , N\}),
\subseteq)$, discussed in Ref.~\cite{Williams2010}. We focus on the bivariate
case for clarity, although the $\Phi$ID formalism developed here can be applied
to any $N$.} Then, the lattice is built using a natural (partial) order
relationship that exists between the elements of
$\mathcal{A}$~\cite{Williams2010}: for $\bm \alpha, \bm \beta \in \mathcal{A}$,
one says that
\begin{equation}
\bm \alpha \preceq \bm
\beta \quad \text{if} \quad  \text{for all } b\in\bm \beta \text{ there exists } a\in\bm\alpha \text{ such that }a
\subset b.    
\end{equation}
The lattice that encodes the relationship $\preceq$ is known as the redundancy
lattice (Fig.~\subref{fig:lattices}{b}), and guides the construction of the
four terms in the PID.

Our first step in building the foundations of $\Phi$ID is to build a
\textit{product lattice} over $\mathcal{A}\times \mathcal{A}$, in order to
extend the notion of redundancy from PID to the case of multiple source and
target variables (here $X_t^1$, $X_t^2$ and $X_{t+1}^1$, $X_{t+1}^2$
respectively). Intuitively, \phiid is the `product' of two complementary
single-target PIDs, one decomposing the information carried by the past about
the future, and the other decomposing the information carried by the future
about the past (Fig.~\subref{fig:lattices}{a}). To formalise this intuition, we
extend Williams and Beer's \cite{Williams2010} notation, and denote sets of
sources and targets using their indices only with an arrow going from past to
future. Hence, the nodes of the product lattice are denoted as $\bm \alpha
\rightarrow \bm \beta$ for $\bm \alpha, \bm \beta \in \mathcal{A}$.

A natural partial ordering relationship can be establish over the product
lattice as follows:
\begin{equation}
\bm \alpha \rightarrow \bm \beta  \preceq \bm \alpha' \rightarrow \bm \beta' 
\quad \text{iff} \quad 
\bm \alpha \preceq \bm \alpha'
\:\: \text{and} \: \:
\bm \beta \preceq \bm \beta'.
\end{equation}
This relationship establishes a lattice structure\footnote{A proof of this is
provided in the Appendix.}, which for the case of a bipartite system consists
of 16 nodes (Fig.~\subref{fig:lattices}{c}).

\subsection*{Redundancies and atoms}

The other ingredient in the PID recipe -- besides the redundancy lattice -- is
a \emph{redundancy function}, $I_\cap$, that quantifies the amount of
`overlapping' information about the target that is common to a set of sources
$\bm\alpha \in \mathcal{A}$~\cite{Williams2010}. The redundancy function in a
PID, $I_\cap^{\bm\alpha}$, encompasses the following terms in the case of two
source variables:
\begin{itemize}
    \item $I_\cap^{\{1\}\{2\}}$ is the information about the target that is in either source, 
    \item $I_\cap^{\{i\}}$ is the information in source $i$, and
    \item $I_\cap^{\{12\}}$ is the information that is in both sources when considered together. 
\end{itemize}
This subsection extends the notion of overlapping information to the
multi-target setting.

For a given $\bm \alpha \rightarrow \bm \beta \in \mathcal{A} \times
\mathcal{A}$, the overlapping information that is common to sources $\bm\alpha$
and can be seen in targets $\bm\beta$ is denoted as $\IR{\bm \alpha}{ \bm \beta
}$ and referred to as the \emph{double-redundancy function}. In the following,
we assume that the double-redundancy function satisfies two axioms:
\begin{itemize}
  \item \textbf{Axiom 1 (compatibility)}: if $\bm \alpha =\{\alpha_1,\dots,\alpha_J\}$
  and $\bm \beta =\{\beta_1,\dots,\beta_K\}$ with %
  $\bm\alpha,\bm\beta\in\mathcal{A}$ and $\alpha_j,\beta_k$ non-empty subsets of $\{1,\dots,N\}$, 
  then the following cases
  can be reduced to the redundancy of %
  PID or the mutual
  information:\footnote{Here we use the shorthand notation $\bm X_t^{\alpha} := (X_t^{i_1},\dots,X_t^{i_K})$ for $\alpha = \{i_1,\dots,i_K\}$.}
\begin{equation}
\IR{\bm \alpha}{\bm \beta} = 
\begin{cases}
  \texttt{Red}(\bm X_t^{\alpha_1},\dots, \bm X_t^{\alpha_J}; \bm X_{t'}^{\beta_1})
&\text{if}\quad K=1,\\
\texttt{Red}(\bm X_{t'}^{\beta_1},\dots, \bm X_{t'}^{\beta_K}; \bm X_t^{\alpha_1})
&\text{if}\quad J=1,\\
I(\bm X_t^{\alpha_1}; \bm X_{t'}^{\beta_1})
&\text{if}\quad J=K=1.
\end{cases}\nonumber
\end{equation}
  \item \textbf{Axiom 2 (partial ordering)}: if $\bm \alpha \rightarrow \bm
  \beta \preceq \bm \alpha' \rightarrow \bm \beta'$ then $\IR{\bm \alpha}{\bm
  \beta} \leq \IR{\bm \alpha'}{\bm \beta'}$.
\end{itemize}
Intuitively, the first axiom guarantees that any double-redundancy function in
$\Phi$ID reduces to a PID-type redudancy function when evaluated in certain
atoms, and the second encapsulates the basic desideratum of the
double-redudancy being in agreement with the partial ordering given by the
product lattice.

By exploiting these two axioms, one can define `atoms' that belong to each of
the nodes via the Moebius inversion formula. Concretely, the $\Phi$ID
\textit{atoms} $\PI{\bm \alpha}{ \bm \beta}$ are defined as the quantities that
guarantee the following condition for all $ \bm \alpha \rightarrow \bm\beta \in
\mathcal{A}\times \mathcal{A}$:
\begin{align}
  \IR{\bm \alpha}{ \bm \beta} = \sum_{\substack{ \bm \alpha' \rightarrow\bm\beta' \preceq \bm\alpha\rightarrow\bm\beta}} \PI{\bm\alpha'}{\bm\beta'} ~.
  \label{eq:mi_decomp}
\end{align}
In other words, $\PI{\bm \alpha}{\bm \beta}$ corresponds to the information
contained in node $\bm\alpha\rightarrow\bm\beta$ and not in any node below it
in the lattice. These are analogues to the redundant, unique, and synergistic
atoms in standard PID, but using the product lattice as a scaffold. By
inverting this relationship, one can find a recursive expression for
calculating $I_{\partial}$ as
\begin{align}\label{eq:moebius}
  \PI{\bm \alpha}{\bm \beta} = \IR{\bm \alpha}{\bm \beta} - \sum_{\substack{ \bm \alpha' \rightarrow\bm\beta' \prec \bm\alpha\rightarrow\bm\beta}} \PI{\bm\alpha'}{\bm\beta'} ~ .
\end{align}
With all the tools at hand, we can deliver the promised decomposition of the
TDMI in terms of atoms of integrated information, as established in the next
definition.

\begin{definition}
The Integrated Information Decomposition ($\Phi\mathrm{ID}$) of a system with Markovian dynamics is the collection of atoms $I_\partial$ defined from the redundancies $I_\cap$ via
Eq.~\eqref{eq:moebius}, which satisfy
\begin{align}
  \mathrm{TDMI} = I(\bm X_t; \bm X_{t'}) = \sum_{\substack{ \bm \alpha,\bm\beta \in \mathcal{A}}} \PI{\bm\alpha}{\bm\beta} ~ .
\end{align}
\end{definition}

In this way, the $\Phi$ID of two time series gives 16 atoms that correspond to
the lattice shown in Figure~\subref{fig:lattices}{c}, which are computed via a
linear transformation over the 16 redundancies. Importantly, Axioms 1 and 2
allow us to compute all the $I_\cap$ terms once a single-target PID redundancy
function $\texttt{Red}(\cdot)$ has been chosen with the sole exception of
$\IR{\{1\}\{2\}}{\{1\}\{2\}}$.\footnote{This can be verified directly by
studying the coefficients of the linear system of equations that relate
redundancies and atoms.} All this is summarised in the following result.

\begin{proposition}[15-for-free]\label{prop:15forfree}
Axioms 1 and 2 provide unique values for the 16 atoms of the product lattice
after one defines (i) a single-target redundancy function
$\texttt{Red}(\cdot)$, and (ii) an expression for
$\PI{\{1\}\{2\}}{\{1\}\{2\}}$.
\end{proposition}
Therefore, in the same way as in PID the definition of $\texttt{Red}(\cdot)$
gives 3 other terms (unique and synergy) as side-product,
Proposition~\ref{prop:15forfree} shows that in $\Phi \text{ID}$ the addition of
the double-redundancy function $\PI{\{1\}\{2\}}{\{1\}\{2\}}$ gives the 15 other
terms for free. In the Supplementary Material we describe \phiid extensions of
two common PID redundancy functions (Ince's CCS~\cite{Ince2017} and Barrett's
MMI~\cite{Barrett2015} measures), which we use for all numerical examples in
this paper.


\section*{Supporting information}

\paragraph*{S1 Appendix.}
\label{S1_Appendix}
{\bf Proofs and technical details.}

\section*{Acknowledgments}

The authors thank Julian Sutherland for valuable discussions. P.A.M.M. was
supported by the Wellcome Trust (grant number 210920/Z/18/Z). F.E.R. was
supported by the Ad Astra Chandaria Foundation, and by the European Union’s
H2020 research and innovation programme under the Marie Sk\l{}odowska-Curie
grant agreement No. 702981. A.I.L. is supported by a Gates Cambridge
Scholarship. A.K.S. and A.B.B. acknowledge support from the Dr Mortimer and
Theresa Sackler Foundation. A.K.S. also acknowledges support from the Canadian
Institute for Advanced Research (CIFAR) Program on Brain, Mind, and
Consciousness.



\bibliography{main}

\end{document}


\title{Supporting Information for\\~\\Towards an extended taxonomy of information dynamics via\\Integrated Information Decomposition}

\author{Pedro A.M. Mediano}
\thanks{P.M. and F.R. contributed equally to this work.\\E-mail: pam83@cam.ac.uk; f.rosas@imperial.ac.uk}
\affiliation{Department of Psychology, University of Cambridge, Cambridge, UK}

\author{Fernando E. Rosas} 
\thanks{P.M. and F.R. contributed equally to this work.\\E-mail: pam83@cam.ac.uk; f.rosas@imperial.ac.uk}
\affiliation{Centre for Psychedelic Research, Imperial College London, London, UK}
\affiliation{Data Science Institute, Imperial College London, London, UK}
\affiliation{Center for Complexity Science, Imperial College London, London, UK}

\author{Andrea I. Luppi}
\affiliation{University Division of Anaesthesia, University of Cambridge, Cambridge, UK}
\affiliation{Department of Clinical Neurosciences, University of Cambridge, Cambridge, UK}
\affiliation{Leverhulme Centre for the Future of Intelligence, University of Cambridge, Cambridge, UK}

\author{\mbox{Robin L. Carhart-Harris}}
\affiliation{Centre for Psychedelic Research, Imperial College London, London, UK}

\author{Daniel Bor}
\affiliation{Department of Psychology, University of Cambridge, Cambridge, UK}

\author{Anil K. Seth}
\affiliation{Sackler Center for Consciousness Science, University of Sussex, Brighton, UK}

\author{Adam B. Barrett}
\affiliation{Sackler Center for Consciousness Science, University of Sussex, Brighton, UK}
\affiliation{The Data Intensive Science Centre, University of Sussex, Brighton, UK}

\date{\today}

\maketitle

\section{The product of two lattices is a lattice}

A lattice is a partially ordered set $(\mathcal{A}, \preceq)$ for which every
pair of elements $a,b$ has a well-defined \textit{meet} $a \wedge b$ and
\textit{join} $a \vee b$, which correspond to their common greatest lower bound
(infimum) and common least upper bound (supremum),
respectively~\cite{charalambides2002enumerative}. Here we prove that, if
$(\mathcal{A},\preceq)$ is a lattice, then the product lattice $(\mathcal{A}
\times \mathcal{A}, \preceq^*)$ equipped with the order relationship
\begin{equation}
\alpha \rightarrow \beta \preceq^* \alpha' \rightarrow \beta' 
\quad \text{if and only if} \quad 
\alpha \preceq \alpha'
\:\: \text{and} \: \:
\beta \preceq \beta',
\end{equation}
\noindent is also a lattice, where $\alpha, \beta, \alpha', \beta' \in
\mathcal{A}$. As a corollary of this, given that the set and partial ordering
relationship used in PID are a lattice \cite{Williams2010,Crampton2001}, then
the set and partial ordering relationship used in $\Phi$ID are also a lattice.

For compactness, let us use the notation $\gamma = \alpha \rightarrow \beta$
and $\gamma' = \alpha' \rightarrow \beta'$ for $\gamma,\gamma'\in
\mathcal{A}\times\mathcal{A}$. To prove the lattice structure of $(\mathcal{A}
\times \mathcal{A}, \preceq^*)$ it suffices to show that

\begin{enumerate}
    \item $\gamma \meet^* \gamma' := \alpha \meet \alpha' \rightarrow \beta \meet \beta'$ is a valid meet; and
    \item $\gamma \join^* \gamma' := \alpha \join \alpha' \rightarrow \beta \join \beta'$ is a valid join.
\end{enumerate}
Note that the fact that $(\mathcal{A},\preceq)$ is a lattice implies that
$\alpha \meet \beta$ and $\alpha \join \beta$ are well-defined for all
$\alpha,\beta\in\mathcal{A}$.

Let us begin with the meet, for which we use $m = \gamma \meet^* \gamma'$ as a
shorthand notation. First, one can directly check that $m \preceq^* \gamma$ and
$m \preceq^* \gamma'$, given the definition of $\preceq^*$ above and the fact
that $\alpha \meet \alpha' \preceq \alpha$ (and similarly for $\alpha'$,
$\beta$, and $\beta'$). Next, we need to prove that for any $\gamma'' =
\alpha'' \rightarrow \beta'' \in \mathcal{A} \times \mathcal{A}$ such that
$\gamma'' \preceq^* \gamma$ and $\gamma'' \preceq^* \gamma'$, we have $\gamma''
\preceq^* m$ (i.e. that $m$ is the greatest lower bound of $\gamma$ and
$\gamma'$). To see this, note that the conditions $\gamma'' \preceq^* \gamma$
and $\gamma'' \preceq^* \gamma'$ imply the following four statements:
\begin{align*}
    \alpha'' & \preceq ~ \alpha ~, \\
    \alpha'' & \preceq ~ \alpha' ~,\\
    \beta''  & \preceq ~ \beta ~,\\
    \beta''  & \preceq ~ \beta' ~ .
\end{align*}
Using these relationships and the $\meet$ operator from $\mathcal{A}$, one can
show that $\alpha'' \preceq \alpha \meet \alpha'$ and $\beta'' \preceq \beta
\meet \beta'$, which in turn implies that $\gamma'' \preceq^* m$. Finally, the
proof for the join is analogous, replacing $\meet$ with $\join$ and $\preceq$
with $\succeq$.

\section{Decomposing PID atoms}

Equation (4) in the main text shows how to decompose redundancies in the
product lattice in terms of $\Phi$ID atoms. Here we provide a more general
statement, that allows us to decompose not only redundancies, but also other
PID atoms. The goal of this appendix is to build stronger connections between
PID and \phiid, and to extend Proposition 1 to allow greater flexibility for
specifying a \phiid function.

Note that the \phiid framework applies to any pair of sets of source and target
variables, which need not correspond to the past and future states of a
dynamical system. To highlight the generality of the \phiid framework, for the
rest of this supplementary material we use the notation $\bX = \{X_1, X_2,
...\}$ to refer to the sources, and analogously $\bY = \{Y_1, Y_2, ...\}$ for
the targets. The expressions in the main text can be recovered by simply
setting $\bX \coloneqq \bX_t, \bY \coloneqq \bX_{t+1}$.

For the forward PID, and borrowing the notation from Williams and Beer
\cite{Williams2010}, given a non-empty set of `future' variables $F \in
\mathcal{P}(\{Y_1, ... , Y_N\})$ and an an element of the redundancy lattice
$\bm\alpha \in \mathcal{A}$, let us denote by $\Pi_F(\bm\alpha; F)$ the $\bm\alpha$ atom
of the PID decomposition for $I(\bX; F)$, such that
\begin{align}
    I(\bX; F) = \sum_{\bm\alpha \in \mathcal{A}} \Pi_F(\bm\alpha; F) ~ .
\end{align}
We use an analogous notation for the backward PID, with a corresponding
non-empty set of `past' variables $P \in \mathcal{P}(\{X_1, ... , X_N\})$ and
$\bm\beta \in \mathcal{A}$, such that
\begin{align}
    I(P; \bY) = \sum_{\bm\beta \in \mathcal{A}} \Pi_B(P; \bm\beta) ~ .
\end{align}

Then, these quantities can be further decomposed in \phiid atoms as
\begin{subequations}
\begin{gather}
    \Pi_F(\bm\alpha; F) = \sum_{\gamma \preceq F} \PI{\bm\alpha}{\gamma} ~ , \label{eq:pid_decomp} \\
    \Pi_B(P;  \bm\beta) = \sum_{\gamma \preceq P} \PI{\gamma}{\bm\beta} ~ .
\end{gather}
\end{subequations}

Note that the sum runs only across one of the sets (instead of both as it does
in Eq.~(4) of the main text), and that every element in $\mathcal{P}(\{1, ... ,
N\})$ is also in $\mathcal{A}$, and hence the partial order relationship in the
sums above is well-defined. As a few examples, in a bivariate system the
following forward PID atoms decompose as:
\begin{align*}
    \texttt{Red}(X_1, X_2; Y_i) &= \Pi_F(\{1\}\{2\}; Y_i) \\
    &= \PI{\{1\}\{2\}}{\{1\}\{2\}} + \PI{\{1\}\{2\}}{\{i\}}~, \\[2ex]
    \texttt{Syn}(X_1, X_2; Y_i) &= \Pi_F(\{12\}; Y_i) \\
                                &= \PI{\{12\}}{\{1\}\{2\}} + \PI{\{12\}}{\{i\}} ~,\\[2ex]
    \texttt{Un}(X_1; Y_1Y_2 | X_2)  &= \Pi_F(\{1\}; Y_1Y_2) \\
                                &= \PI{\{1\}}{\{1\}\{2\}} + \PI{\{1\}}{\{1\}} \\
                                &~ ~ ~+ \PI{\{1\}}{\{2\}} + \PI{\{1\}}{\{12\}}~.
\end{align*}

These decompositions can be used to prove Proposition~1 of the main text.
Adopting a view of $\Phi$ID as a linear system of equations, one needs 16
independent equations to solve for the 16 unknowns that are the \phiid atoms.
Of those, 9 are given by standard Shannon mutual information (specifically,
$I(X_i;Y_j)$, $I(X_1X_2; Y_i)$, $I(Y_1Y_2; X_i)$, and $I(X_1X_2; Y_1Y_2)$, for
$i,j=\{1,2\}$) decomposed with Eq.~(4) of the main text, and 6 are given by the
single-target PIDs ($\texttt{Red}(X_1, X_2; Y_1)$, $\texttt{Red}(X_1, X_2;
Y_2)$, and $\texttt{Red}(X_1, X_2; Y_1Y_2)$, as well as the 3 corresponding
backward PIDs) decomposed by the expression above. Finally, one only need to
add one individual \phiid atom to make the 16 equations needed, and the system
can be solved for all other atoms.

Taking these results together, Proposition~1 in the main text can be
generalised as follows: a valid $\Phi$ID can be defined not only in terms of
redundancy, but also in terms of unique information or synergy. This is
equivalent to the case of PID, for which decompositions based on unique
information \cite{James2018} or synergy
~\cite{quax2017quantifying,rosas2020operational} have been proposed.

\section{Computing the $\Phi$ID atoms}
\label{sec:computing}

Computation of \phiid proceeds following the same general steps as in PID:
first the intersection information $\IR{\bm \alpha}{\bm \beta}$ is computed for every
node; and then the integrated information atoms $\PI{\bm \alpha}{\bm \beta}$ are
obtained as the solution to a linear system of equations representing the Moebius
inversion.

To compute the double-redundancy atom, $\PI{\{1\}\{2\}}{\{1\}\{2\}}$, for
numerical applications, we assume all systems are distributed as a multivariate
Gaussian distribution, and use a \phiid extension of Ince's \emph{Common
Change in Surprisal} (CCS) redundancy function~\cite{Ince2017}. As per the
compatibility axiom, we formulate a multi-target CCS function that reduces to
the original when only a single target is specified.

In line with Ince~\cite{Ince2017}, we define
$I^{\{1\}\{2\}\to\{1\}\{2\}}_{\partial,\mathrm{CCS}}$ using pointwise (or
\emph{local}) information measures~\footnote{For the rationale behind and
extended discussion of local information measures see
Lizier~\cite{Lizier2010}.}. As a first step, we use the inclusion-exclusion
principle to formulate a local `multi-target co-information' $c(\bm x; \bm y)$,
defined as
\begin{align}
c(\bm x; \bm y) \coloneqq \sum_{\bm\alpha\to\bm\beta \in \overline{\mathcal{A}^2}} (-1)^{f(\bm\alpha, \bm\beta)+1} ~ i_\cap^{\bm\alpha\to\bm\beta}(\bm x; \bm y) ~ ,
\end{align}
where $i_\cap^{\bm\alpha\to\bm\beta}(\bm x; \bm y)$ is the pointwise redundancy
function, $\overline{\mathcal{A}^2}$ is the set of nodes in the product lattice
excluding the lowest node, and $f(\bm\alpha, \bm\beta) = \sum_{a \in \bm\alpha}
|a| + \sum_{b \in \bm\beta} |b|$~\footnote{Equivalently, $f(\bm\alpha,
\bm\beta)$ represents the length of the shortest path in the product lattice
between the node $\bm\alpha\to\bm\beta$ and the lowest node}. Note that
$i_\cap^{\bm\alpha\to\bm\beta}(\bm x; \bm y)$ above corresponds to the standard
pointwise mutual information and a single-target PID redundancy function, which
we take here to be the usual CCS function as defined by Ince~\cite{Ince2017}.
For the bivariate \phiid, a formal calculation shows that
\begin{align*}
c(\bm x; \bm y) = i_\partial^{\{1\}\{2\}\to\{1\}\{2\}}(\bm x; \bm y) - i_\partial^{\{12\}\to\{12\}}(\bm x; \bm y) ~ .
\end{align*}
Please note that, as for the co-information in the standard PID case, $c(\bm x;
\bm y)$ is a `whole-minus-sum' measure\cite{Rosas2019} that can be computed
without any \phiid atoms explicitly. Then, as the second step in the
definition, given a large set of $M$ samples $\{\bm x^{(i)}, \bm
y^{(i)}\}_{i=1}^M$ we define the set $\mathcal{S} \subseteq \{1, ..., M\}$ as
the subset of samples for which all marginal pointwise mutual informations, as
well as the pointwise full mutual information $i(\bm x; \bm y)$, have the same
sign. With this, we are finally able to define the Gaussian CCS
double-redundancy function.

\begin{definition}
\textbf{Double-redundancy based on common change in surprisal}. For a given set
of variables $(\bm X,\bm Y)$, the double-redundancy based on common change in
surprisal is defined as
\begin{equation}
I^{\{1\}\{2\}\to\{1\}\{2\}}_{\partial,\mathrm{CCS}} \coloneqq \sum_{i \in \mathcal{S}} c(\bm x^{(i)}; \bm y^{(i)})
\end{equation}
\end{definition}

To show the presented results do not depend on the specific choice of
the CCS function, we also formulate a \phiid extension of the
\emph{Minimum Mutual Information} (MMI)
PID~\cite{Barrett2015}:

\begin{definition}
\textbf{Double-redundancy based on minimum mutual information}. For a given set of
variables $(\bm X,\bm Y)$, the double-redundancy
based on minimum mutual information is defined as
\begin{equation}
I^{\{1\}\{2\}\to\{1\}\{2\}}_{\partial,\mathrm{MMI}} \coloneqq \min_{i,j} I(X_i; Y_j).
\end{equation}
\end{definition}

In both cases (CCS and MMI), it is direct to check the proposed definitions
satisfy the compatibility axiom with respect to their original PID
definitions~\cite{Ince2017,Barrett2015}. At the same time, and although the
extensions presented here seem the most natural, they are not the only possible
ones that are compatible with the originals, and in principle any function that
satisfies the double-redundancy axioms can be used to compute \phiid.

\section{Results of section `Different types of integration'}

Here we present calculations for the example systems in Fig. 4 of the main
text. These proofs hold for all $\Phi$ID that satisfy the partial ordering
axiom of $\IR{\bm\alpha}{\bm\beta}$ (Axiom 2 in the main text), have a non-negative
double-redundancy function $I^{\{1\}\{2\}\to\{1\}\{2\}}\geq 0$, and satisfy the
following bound that follows from the basic properties of PID
(c.f.~\cite{rosas2016understanding}):
\begin{align}
  \texttt{Red}(X,Y;Z) \leq \min\{ I(X;Z), I(Y;Z) \} ~ .
\end{align}
Let us examine the three systems in turn:

\begin{itemize}

\item For the copy transfer system, $Y_2=X_1$, while $X_2$ and $Y_2$ are
independent i.i.d. fair coin flips. Since $Y_2$ is independent from the rest of
the system, $\texttt{Red}(X_1,X_2;Y_2) = \texttt{Red}(X_1,X_2;Y_2) = 0$, and
due to partial ordering $\IR{\{1\}\{2\}}{\{1\}\{2\}} = 0$. Finally, using the
Moebius inversion formula it follows that $\PI{\{1\}}{\{2\}} = I(X_1;Y_2) = 1$
and all other atoms are zero.

\item In the downward XOR system, $X_1$ and $X_2$ are i.i.d. fair coin flips,
$Y_1 = X_1 \oplus X_2$, and $Y_2$ is independent of the rest. Then, it is clear
that $I(X_1,X_2;Y_1,Y_2) = I(X_1,X_2;Y_1) = 1$, while
$I(X_1;Y_1)=I(X_2;Y_1)=0$. Additionally, note that $\IR{\{12\}}{\{1\}\{2\}}=0$,
since $\texttt{Red}(Y_1,Y_2;X_1 X_2) \leq I(Y_2; X_1 X_2) = 0$. All this
implies that all the redundancies (and hence all the atoms) below
$\{12\}\to\{1\}$ are zero, and hence $\PI{\{12\}}{\{1\}}=1$ due to the Moebius
inversion formula.

\item Finally, consider the PPR system where $X_1,X_2,Y_1$ are i.i.d. fair coin
flips and $Y_2$ is such that $X_1\oplus X_2 = Y_1\oplus Y_2$. Then
$I(X_1,X_2;Y_1) = I(X_1,X_2;Y_2) = I(X_1;Y_1,Y_2)=I(X_2;Y_1,Y_2)=0$. This
implies that all redundancies (and hence atoms) except $\IR{\{12\}}{\{12\}}$
are zero, and hence using again the Moebius inversion formula
$\PI{\{12\}}{\{12\}}=I(X_1,X_2;Y_1,Y_2)=1$.

\end{itemize}

\section{Results related to measures of integrated information}

In this appendix we prove the results in Table 1 of the main text, that shows
whether each of four measures of integrated information ($\wmsphi$, CD, $\psi$,
$\Phi_G$) are positive, negative, or zero in a system containing only one
\phiid atom. A succinct definition of each measure is given below, and a
comprehensive review and comparison of these and other measures can be found in
Ref.~\cite{Mediano2019}.

Throughout this section we focus on bivariate systems, and use $i,j$ as
variable indices, with $i \neq j$. To complete the proof we will first show
that it is possible to build systems with exactly one bit of information in one
\phiid atom, and we will then compute the four measures on those systems.

Let us begin with the design of systems with one specific \phiid atom.
Intuitively, this can be accomplished with a suitable combination of COPY and
XOR gates for redundant and synergistic sets of variables, respectively. More
formally, the procedure to build a system with $\PI{\bm\alpha}{\bm\beta} = 1$ and all
other atoms equal to zero is as follows:

\begin{enumerate}
    \item Sample $w$ from a Bernoulli distribution with $p = 0.5$.
    \item Sample $\bm x$ based on $\bm\alpha$:
\begin{itemize}
    \item If $\bm\alpha = \{1\}\{2\}$, then $x_1 = x_2 = w$.
    \item If $\bm\alpha = \{i\}$, then $x_i = w$ and $x_j$ is sampled from a Bernoulli distribution with $p = 0.5$.
    \item If $\bm\alpha = \{12\}$, then $\bm x$ is a random string with parity $w$.
\end{itemize}
    \item Sample $\bm y$ based on $\bm\beta$ analogously.
\end{enumerate}

In all cases there will be one bit of information ($w$) shared between $\bX$
and $\bY$, hence $I(\bX; \bY) = 1$ for any choice of $\bm\alpha, \bm\beta$. This can
be proven using the fact that for any $\bm\alpha, \bm\beta$, one has $H(W) = 1$, $H(W
| \bm X) = H(W | \bm Y) = 0$, and $p(\bm x, \bm y, w) = p(\bm x|w)p(\bm
y|w)p(w)$. To do so, let us start from the mutual information chain rule:
\begin{align*}
    I(\bm X; \bm YW) &= I(\bm X; W) + I(\bm X; \bm Y|W) \\
             &= I(\bm X; \bm Y) + I(\bm X; W|\bm Y) ~ .
\end{align*}
Rearranging the above terms, one can find that
\begin{align*}
    I(\bm X; \bm Y) = I(\bm X; W) + I(\bm X; \bm Y|W) - I(\bm X; W|\bm Y) ~ ,
\end{align*}
where $I(\bm X; W) = H(W) - H(W|\bm X) = 1$ and $I(\bm X; \bm Y|W) = 0$.
Finally, one finds that
\begin{align*}
    I(\bm X; W|\bm Y) &= H(\bm X|\bm Y) + H(W|\bm Y) - H(\bm XW|\bm Y) \\
    &= H(\bm X|\bm Y) + H(W|\bm Y) -  \big[ H(\bm X|\bm Y) + H(W|\bm X\bm Y) \big] \\
    &= 0 ~ ,
\end{align*}
which concludes the proof that $I(\bm X; \bm Y) = 1$. Furthermore, following a
procedure similar to those in the previous section, it can be shown that any
\phiid that satisfies the axioms described above (partial ordering,
non-negative double-redundancy, and upper-bounded redundancy) correctly assigns
1 bit of information to $\PI{\bm\alpha}{\bm\beta}$, and 0 to all other atoms.

Now that we have built these 16 single-atom systems, let us move to the
integration measures of interest. For CD, $\psi$, and \wmsphi, we will proceed
by decomposing them in terms of \phiid atoms and checking whether each atom is
positive (+), negative (--), or absent (0) from the decomposition to obtain the
results in Table 1 of the article. Let us begin with CD, defined as the sum of
transfer entropies from one variable to the other:
\begin{align}
\begin{split}
    \mathrm{CD} &= \frac{1}{2} \sum_{i=1}^2 I(X_i; Y_j | X_j) \\
                &= \frac{1}{2} \sum_{i=1}^2 \Big( \PI{\{i\}}{\{1\}\{2\}} + \PI{\{i\}}{\{j\}} + \PI{\{12\}}{\{1\}\{2\}} + \PI{\{12\}}{\{j\}} \Big) ~ .
\end{split}
\end{align}
Similarly, for $\psi$ the atoms can be extracted from the decomposition of
$\texttt{Syn}(X_1, X_2; Y_1 Y_2)$ in Eq.~\eqref{eq:pid_decomp}:
\begin{align}
\begin{split}
    \psi &= \texttt{Syn}(X_1, X_2; Y_1 Y_2) \\ &= \PI{\{12\}}{\{1\}\{2\}} + \PI{\{12\}}{\{1\}} + \PI{\{12\}}{\{2\}} + \PI{\{12\}}{\{12\}} ~ .
\end{split}
\end{align}
For \wmsphi, the atoms can be extracted from the decomposition of Eq.~(9) in
the main text:
\begin{align}
\begin{split}
    \wmsphi = & -\PI{\{1\}\{2\}}{\{1\}\{2\}} + \PI{\{1\}\{2\}}{\{12\}} + \psi + \sum_{i=1}^2 \left( \PI{\{i\}}{\{j\}} + \PI{\{i\}}{\{12\}} \right) ~ .
\end{split}
\end{align}

The $\Phi_G$ case is slightly more involved, since it is not easily
decomposable into a sum of \phiid atoms. According to the definition of
$\Phi_G$ \cite{Oizumi2016}, for a system given by the joint probability
distribution $p(\bX, \bY)$ one has
\begin{align*}
    \Phi_G = \min_{q \in \mathcal{M}_G} D_{\mathrm{KL}}(p \| q) ~ ,
\end{align*}
where $\mathcal{M}_G$ is the manifold of probability distributions that satisfy
the constraints
\begin{align}
    q(Y_i | \bX) = q(Y_i | X_i) ~ .
    \label{eq:phig_constraint}%
\end{align}
Therefore, it suffices to check whether the probability distribution of the
system satisfies the constraints in Eq.~\eqref{eq:phig_constraint} --- if it
does, then $\Phi_G = 0$, and otherwise $\Phi_G > 0$ ---, which can be easily
verified for each system separately to obtain the $\Phi_G$ column in Table 1,
concluding the proof.

\vspace{20pt}

\section{Results of section `Why whole-minus-sum $\Phi$ can be negative'}

In this appendix we describe the details of the noisy autoregressive system and how to
compute its \phiid to yield the results shown in Figure 5 of the main text.

Given the past state of the system $x_t^1,x_t^2$, the next state is given by
\begin{align*}
    x_{t+1}^1 &= a(x_t^1 + x_t^2) + \varepsilon_{t+1}^1 \\
    x_{t+1}^2 &= a(x_t^1 + x_t^2) + \varepsilon_{t+1}^2 ~ ,
\end{align*}
where $a=0.4$ is a fixed coupling parameter and $\varepsilon_t^1,
\varepsilon_t^2$ are zero-mean unit-variance white noise processes with
correlation $c$. All information-theoretic functionals are computed with
respect to the system's stationary distribution, which can be shown to be a
Gaussian and analytically calculated by means of a discrete Lyapunov
equation~\cite{Mediano2019,Barrett2011}. Once this distribution is obtained,
the values of the atoms can be obtained following the procedures in
Sec.~\ref{sec:computing} above.

\section{Simulation and analysis of whole-brain computational model}

\begin{table}
  \caption{Dynamic Mean Field (DMF) model parameters}
  \centering
  \def\arraystretch{1.2}
  \begin{tabular}{ | l | l | l | p{5cm} |}
    \hline
    \textit{Parameter} & \textit{Symbol} & \textit{Value}  \\ \hline\hline
    External current & $I_0$ & 0.382 nA \\ \hline
    Excitatory scaling factor for $I_0$ & $W_\text{E}$ & 1   \\ \hline
    Inhibitory scaling factor for $I_0$ & $W_\text{I}$ & 0.7 \\ \hline
    Local excitatory recurrence & $w_+$ &  1.4\\ \hline
    Excitatory synaptic coupling & $J_\mathrm{NMDA}$ & 0.15 nA \\ \hline
    Threshold for $F(I_n^\text{(E)})$ & $I_\text{thr}^\text{(E)}$ &  0.403 nA\\ \hline
    Threshold for $F(I_n^\text{(I)})$  & $I_\text{thr}^\text{(I)}$ &  0.288 nA\\ \hline
    Gain factor of $F(I_n^\text{(E)})$ & $g_\text{E}$ & 310 nC$^{-1}$  \\ \hline
    Gain factor of $F(I_n^\text{(I)})$ & $g_\text{I}$ & 615 nC$^{-1}$ \\ \hline
    Shape of $F(I_n^\text{(E)})$ around $I_\text{thr}^\text{(E)}$ & $d_\text{E}$ &  0.16 s \\ \hline
    Shape of $F(I_n^\text{(I)})$ around $I_\text{thr}^\text{(I)}$ & $d_\text{I}$ &  0.087 s\\ \hline
    Excitatory kinetic parameter & $\gamma$ & 0.641 \\ \hline
    Amplitude of uncorrelated Gaussian noise $v_n$ &$\sigma$ &  0.01 nA\\ \hline
    Time constant of NMDA & $\tau_\mathrm{NMDA}$ &  100 ms\\ \hline
    Time constant of GABA & $\tau_\mathrm{GABA}$ &  10 ms\\ \hline
  \end{tabular}
  \label{tab:params}
\end{table}

To explore information decomposition in realistic neurophysiological data we
study the Dynamic Mean-Field (DMF) model by Deco \textit{et
al.}~\cite{Deco2014,Deco2018}, which consists of a set of coupled differential
equations modelling the average activity of multiple interacting brain regions.
These equations represent each brain region as two reciprocally coupled
neuronal populations, one excitatory and one inhibitory, with the corresponding
synaptic currents $I^\text{(E)}$ and $I^\text{(I)}$ are mediated by NMDA and
GABA\textsubscript{A} receptors respectively. Different brain regions are
coupled via their excitatory populations only, and the structural connectivity
is given by the matrix $C$. The structural connectivity matrix was obtained
from the HCP 900 subjects data release~\cite{VanEssen2013,Glasser2013}, and was
preprocessed in the same was as in Ref.~\cite{Luppi2020}, resulting in an
83$\times$83 connectivity matrix corresponding to the Lausanne-83 brain
parcellation~\cite{Cammoun2012}. For all other aspects of model configuration
and simulation we follow Herzog \emph{et al.}~\cite{Herzog2020}, and reproduce
all relevant details here for convenience.

The full model is given by
\begin{align*}
  I_j^\text{(E)} &= W_\text{E} I_0 + w_+ J_{\mathrm{NMDA}} S_j^\text{(E)} + G J_{\mathrm{NMDA}} \sum_{k=1}^N C_{jk} S_p^\text{(E)} - J^{\mathrm{FIC}}_j S_j^\text{(I)} \\
  I_j^\text{(I)} &= W_\text{I} I_0 + J_{\mathrm{NMDA}} S_j^\text{(E)} - S_j^\text{(I)} \\
r_j^\text{(E)} &= F\left(I_j^\text{(E)}\right) = \frac{g_\text{E} \left( I_j^\text{(E)} - I_\text{thr}^\text{(E)} \right)} { 1 - \exp\left(-d_\text{E}\; g_\text{E} \left( I_j^\text{(E)} - I_\text{thr}^\text{(E)} \right) \right)} \\
r_j^\text{(I)} &= F\left(I_j^\text{(I)}\right) = \frac{g_\text{I} \left( I_j^\text{(I)} - I_\text{thr}^\text{(I)} \right)} { 1 - \exp\left(-d_\text{I}\;  g_\text{I} \left( I_j^\text{(I)} - I_\text{thr}^\text{(I)} \right) \right)} \\
\frac{dS_j^\text{(E)}(t)}{dt} &= - \frac{S_j^\text{(E)}}{\tau_{\mathrm{NMDA}}} + \left(1-S_j^\text{(E)}\right) \gamma r_j^\text{(E)} + \sigma v_j(t) \\
\frac{dS_j^\text{(I)}(t)}{dt} &= - \frac{S_j^\text{(I)}}{\tau_{\mathrm{GABA}}} + r_j^\text{(I)} + \sigma v_j(t)
\end{align*}

Above, $j,k$ are indices that run across all $N$ brain regions; $F$ is the
\emph{F-I} curve relating input current to output firing rate of a neural
population; $J^{\mathrm{FIC}}$ is the feedback inhibitory control parameter,
optimised to yield average firing rates of approximately 3 Hz; and the sub- and
superscripts $\text{E}$/$\text{I}$ denote excitatory/inhibitory quantities, respectively. The
model was simulated using a standard Euler-Maruyama integration
method~\cite{Kloeden2013}, using the parameter values shown in
Table~\ref{tab:params}. Note that all parameter values are fixed except the
global coupling $G$, which we vary across simulations. Finally, the
distributions of the simulated BOLD signals are approximated via Gaussian
distributions, and the procedures in Sec.~\ref{sec:computing} above are applied
to obtain the values of all \phiid atoms.

\section{Numerical results replicated with alternative \phiid measures}

\begin{figure}[ht]
  \centering
  \includegraphics{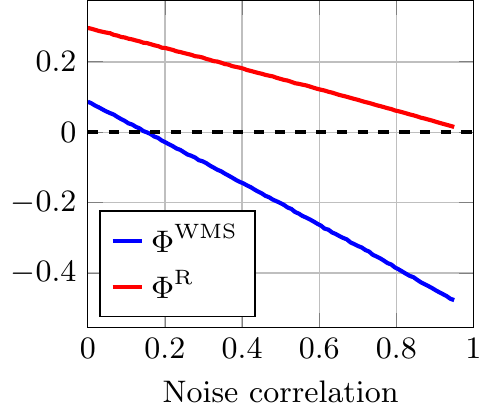}
  \caption{\textbf{Results in two-node AR system replicated with MMI double-redundancy}.}
  \label{fig:ar_supp}
\end{figure}

\begin{figure}[ht]
  \centering
  \includegraphics{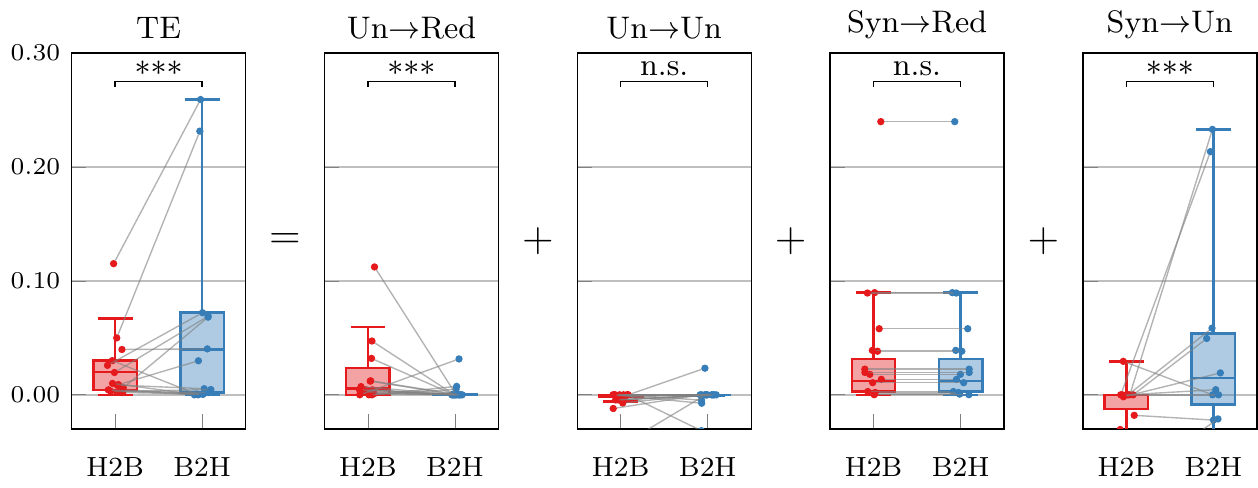}
  \caption{\textbf{Results in Fantasia dataset replicated with MMI double-redundancy}.}
  \label{fig:fantasia_supp}
\end{figure}

\begin{figure}[ht]
  \centering
  \includegraphics{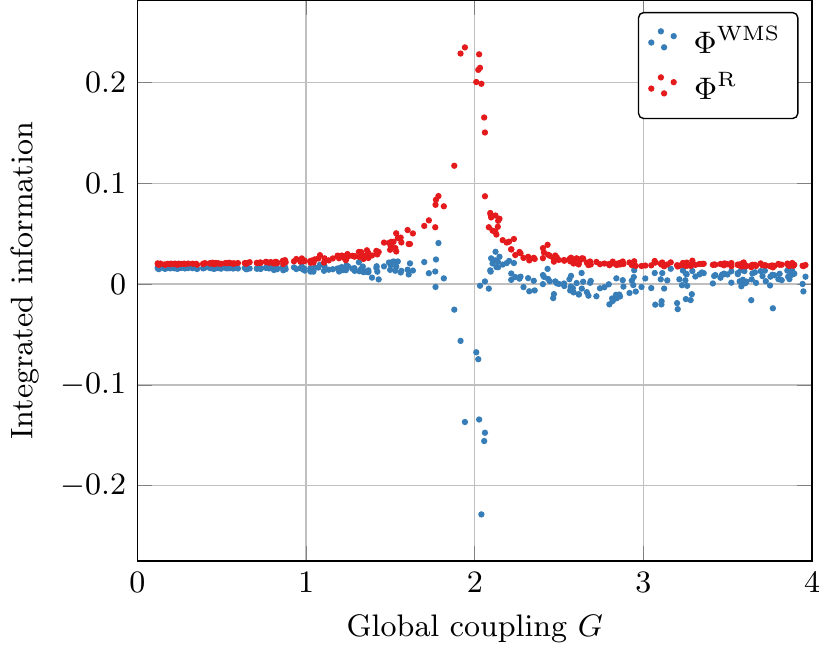}
  \caption{\textbf{Results in DMF whole-brain simulation replicated with MMI double-redundancy}.}
  \label{fig:dmf_supp}
\end{figure}

\bibliography{main.bib}